\documentclass[acmtog]{acmart}

\usepackage{booktabs} 

\usepackage{bm}
\usepackage{ulem}
\usepackage{color}
\usepackage[ruled, linesnumbered]{algorithm2e} 

\SetAlFnt{\small}
\SetAlCapFnt{\small}
\SetAlCapNameFnt{\small}
\SetAlCapHSkip{3pt}
\usepackage{mystyle}
\usepackage{subcaption}
\graphicspath{{./figures/}{./}}
\acmJournal{TOG}
\acmYear{2020}
\acmVolume{39}
\acmNumber{4}
\acmArticle{1}
\acmMonth{7}
\acmDOI{10.1145/3386569.3392407}

\setcopyright{acmcopyright}

\begin{document}

\title{Informative Scene Decomposition for Crowd Analysis, Comparison and Simulation Guidance}

\author{Feixiang He}
\affiliation{%
\institution{University of Leeds}
\department{School of Computing}
\country{United Kingdom}}
\email{fxhe1992@gmail.com}

\author{Yuanhang Xiang}
\affiliation{%
  \institution{Xi'an Jiaotong University}
  \department{School of Computer Science and Technology}
  \country{China}
}
\email{xiangyuanhang@icloud.com}
\author{Xi Zhao}
\authornote{Corresponding author}
\affiliation{
 \institution{Xi'an Jiaotong University}
 \department{School of Computer Science and Technology}
 \country{China}}
\email{zhaoxi.jade@gmail.com}
\author{He Wang}
\authornote{Corresponding author}
\affiliation{%
  \institution{University of Leeds}
  \department{School of Computing}
  \country{United Kingdom}
}
\email{realcrane@gmail.com}

\begin{abstract}
Crowd simulation is a central topic in several fields including graphics. To achieve high-fidelity simulations, data has been increasingly relied upon for analysis and simulation guidance. However, the information in real-world data is often noisy, mixed and unstructured, making it difficult for effective analysis, therefore has not been fully utilized. With the fast-growing volume of crowd data, such a bottleneck needs to be addressed. In this paper, we propose a new framework which comprehensively tackles this problem. It centers at an unsupervised method for analysis. The method takes as input raw and noisy data with highly mixed multi-dimensional (space, time and dynamics) information, and automatically structure it by learning the correlations among these dimensions. The dimensions together with their correlations fully describe the scene semantics which consists of recurring activity patterns in a scene, manifested as space flows with temporal and dynamics profiles. The effectiveness and robustness of the analysis have been tested on datasets with great variations in volume, duration, environment and crowd dynamics. Based on the analysis, new methods for data visualization, simulation evaluation and simulation guidance are also proposed. Together, our framework establishes a highly automated pipeline from raw data to crowd analysis, comparison and simulation guidance. Extensive experiments and evaluations have been conducted to show the flexibility, versatility and intuitiveness of our framework.
\end{abstract}


%
%


\begin{teaserfigure}
    \centering
    \includegraphics[width=1.\textwidth]{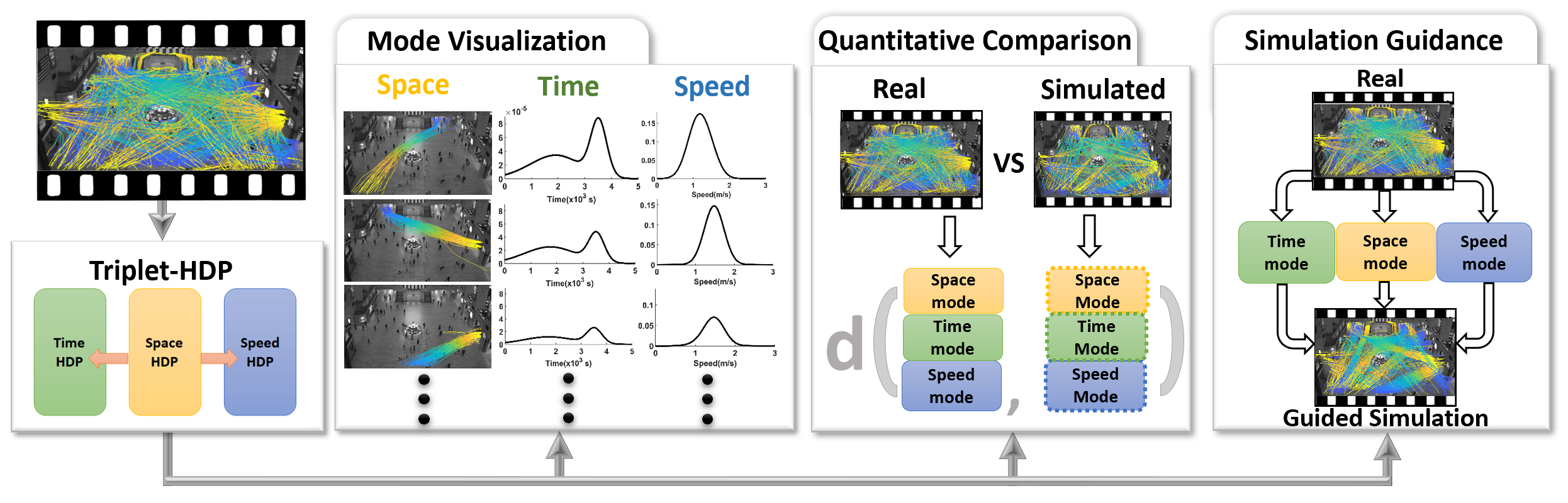}
    \caption{Overview of our framework.}
    \label{fig:teaser}
\end{teaserfigure}

\maketitle

\section{Introduction}
Crowd simulation has been intensively used in computer animation, as well as other fields such as architectural design and crowd management. The fidelity or realism of simulation has been a long-standing problem. The main complexity arises from its multifaceted nature. It could mean high-level global behaviors \cite{narain_aggregate_2009}, mid-level flow information \cite{wang_path_2016} or low-level individual motions \cite{guy_statistical_2012}. It could also mean perceived realism \cite{Ennis_perceptual_2011} or numerical accuracy \cite{wang_trending_2017}. In any case, analyzing real-world data is inevitable for evaluating and guiding simulations.

The main challenges in utilizing real-world data are data complexity, intrinsic motion randomness and the shear volume. The data complexity makes structured analysis difficult. As the most prevalent form of crowd data, trajectories extracted from sensors contain rich but mixed and unstructured information of space, time and dynamics. Although high-level statistics such as density can be used for analysis, they are not well defined and cannot give structural insights \cite{wang_trending_2017}. Second, trajectories show intrinsic randomness of individual motions \cite{guy_statistical_2012}. The randomness shows heterogeneity between different individuals and groups, and is influenced by internal factors such as state of mind and external factors such as collision avoidance. Hence a single representation is not likely to be able to capture all randomness for all people in a scene. This makes it difficult to guide simulation without systematically considering the randomness. Lastly, with more recording devices being installed and data being shared, the shear volume of data in both space and time, with excessive noise, requires efficient and robust analysis. 

Existing methods that use real-world data for purposes such as qualitative and quantitative comparisons \cite{wang_path_2016}, simulation guidance \cite{Ren_heter_2018} or steering \cite{Lopez_character_2019}, mainly focus on one aspect of data, e.g. space, time or dynamics, and tend to ignore the structural correlations between them. Also during simulation and analysis, motion randomness is often ignored or uniformly modelled  for all trajectories \cite{Helbing_social_1995, guy_statistical_2012}. Ignoring the randomness (e.g. only assuming the least-effort principle) makes simulated agents to walk in straight lines whenever possible, which is rarely observed in real-world data; uniformly modelling the randomness fails to capture the heterogeneity of the data. Besides, most existing methods are not designed to deal with massive data with excessive noise. Many of them require the full trajectories to be available \cite{wolinski_parameter_2014} which cannot be guaranteed in real world, and do not handle data at the scale of tens of thousands of people and several days long.

In this paper, we propose a new framework that addresses the three aforementioned challenges. This framework is centered at an analysis method which automatically decomposes a crowd scene of a large number of trajectories into a series of \textit{modes}. Each mode comprehensively captures a unique pattern of spatial, temporal and dynamics information. Spatially, a mode represents a pedestrian flow which connects subspaces with specific functionalities, e.g. entrance, exit, information desk, etc.; temporally it captures when this flow appears, crescendos, wanes and disappears; dynamically it reveals the speed preferences on this flow. With space, time and dynamics information, each mode represents a unique recurring activity and all modes together describe the \textit{scene semantics}. These modes serve as a highly flexible visualization tool for general and task-specific analysis. Next, they form a natural basis where explicable evaluation metrics can be derived for quantitatively comparing simulated and real crowds, both holistically and dimension-specific (space, time and dynamics). Lastly, they can easily automate simulation guidance, especially in capturing the heterogeneous motion randomness in the data. 

The analysis is done by a new \textit{unsupervised} clustering method based on non-parametric Bayesian models, because manual labelling would be extremely laborious. Specifically, Hierarchical Dirichlet Processes (HDP) are used to disentangle the spatial, temporal and dynamics information. Our model consists of three intertwined HDPs and is thus named Triplet HDPs (THDP). The outcome is a (potentially infinite) number of modes with weights. Spatially, each mode is a crowd flow represented by trajectories sharing spatial similarities. Temporally, it is a distribution of when the flow appears, crescendos, peaks, wanes and disappears. Dynamically, it shows the speed distribution of the flow. The whole data is then represented by a weighted combination of all modes. Besides, the power of THDP comes with an increased model complexity, which brings challenges on inference. We therefore propose a new method based on Markov Chain Monte Carlo (MCMC). The method is a major generalization of the Chinese Restaurant Franchise (CRF) method, which was originally developed for HDP. We refer to the new inference method as Chinese Restaurant Franchise League (CRFL). THDP and CRFL are general and effective on datasets with great spatial, temporal and dynamics variations. They provide a versatile base for new methods for visualization, simulation evaluation and simulation guidance. 

Formally, we propose the first, to our best knowledge, multi-purpose framework for crowd analysis, visualization, simulation evaluation and simulation guidance, which includes:
\begin{enumerate}
    \item a new activity analysis method by unsupervised clustering.
    \item a new visualization tool for highly complex crowd data.
    \item a set of new metrics for comparing simulated and real crowds.
    \item a new approach for automated simulation guidance.
\end{enumerate}

To this end, we have technical contributions which include:
\begin{enumerate}
    \item the first, to our best knowledge, non-parametric method that holistically considers space, time and dynamics for crowd analysis, simulation evaluation and simulation guidance.
    \item a new Markov Chain Monte Carlo method which achieves effective inference on intertwined HDPs.
\end{enumerate}

\section{Related Work}
\subsection{Crowd Simulation}
Empirical modelling and data-driven methods have been the two mainstreams in simulation.  Empirical modelling dominates early research, where observations of crowd motions are abstracted into mathematical equations and deterministic systems. Crowds can be modelled as fields or flows \cite{narain_aggregate_2009}, or as particle systems \cite{Helbing_social_1995}, or by velocity and geometric optimization \cite{Berg_Reciprocal_2008}. Social behaviors including queuing and grouping \cite{lemercier_realistic_2012,Ren_group_2016} have also been pursued. On the other hand, data-driven simulation has also been explored, in using e.g. first-person vision to guide steering behaviors \cite{Lopez_character_2019} or trajectories to extract features to describe motions \cite{lee2007group,karamouzas_crowd_2019}. Our research is highly complementary to simulation research in providing analysis, guidance and evaluation metrics. It aims to work with existing steering and global planning methods.

\subsection{Crowd Analysis}
Crowd analysis has been a trendy topic in computer vision \cite{Wang_Globally_2016,wang_trajectory_2008}. They aim to learn structured latent patterns in data, similar to our analysis method. However, they only consider limited information (e.g. space only or space/time) compared to our method because our method explicitly models space, time, dynamics and their correlations. In contrast, another way of scene analysis is to focus on the anomalies \cite{charalambous2014data}. Their perspective is different from ours and therefore complementary to our approach. Trajectory analysis also plays an important role in modern sports analysis \cite{sha2017fine,sha2018interactive}, but they do not deal with a large number of trajectories as our method does. Recently, deep learning has been used for crowd analysis in trajectory prediction \cite{Xu_encoding_2018}, people counting \cite{Wang_learning_2019}, scene understanding \cite{Liu_ADCrowdNet_2019} and anomaly detection \cite{Sabokrou_deep_2019}.  However, they either do not model low-level behaviors or can only do short-horizon prediction (seconds). Our research is orthogonal to theirs by focusing on the analysis and its applications in simulations.

Besides computer vision, crowd analysis has also been investigated in physics. In \cite{ali2007lagrangian}, Lagrangian Particle Dynamics is exploited for the segmentation of high-density crowd flows and detection of flow instabilities, where the target was similar to our analysis. But they only consider space when separating flows, while our research explicitly models more comprehensive information, including space, time and dynamic. Physics-inspired approaches have also been applied in abnormal trajectory detection for surveillance \cite{mehran2009abnormal,chaker2017social}. An approach based on social force model \cite{mehran2009abnormal} is introduced to describe individual movement in microscopic by placing a grid particle over the image. A local and global social network are built by constructing a set of spatio-temporal cuboids in \cite{chaker2017social} to detect anomalies. Compared with these methods, our anomaly detection is more informative and versatile in providing what attributes contribute to the abnormality.

\subsection{Simulation Evaluation}
How to evaluate simulations is a long-standing problem. One major approach is to compare simulated and real crowds. There are qualitative and quantitative methods. Qualitative methods include visual comparison \cite{lemercier_realistic_2012} and perceptual experiments \cite{Ennis_perceptual_2011}. Quantitative methods fall into model-based methods \cite{golas_hybrid_2013} and data-driven methods \cite{lerner2009data,guy_statistical_2012,wang_path_2016,wang_trending_2017}. Individual behaviors can be directly compared between simulation and reference data \cite{lerner2009data}. However, it requires full trajectories to be available which is difficult in practice. Our comparison is based on the latent behavioral patterns instead of individual behaviors and does not require full trajectories. The methods in \cite{wang_path_2016,wang_trending_2017} are similar to ours where only space is considered. In contrast, our approach is more comprehensive by considering space, time and dynamics. Different combinations of these factors result in different metrics focusing on comparing different aspects of the data. The comparisons can be spatially focused or temporally focused. They can also be comparing general situations or specific modes. Overall, our method provides greater flexibility and more intuitive results.

\subsection{Simulation Guidance}
Quantitative simulation guidance has been investigated before, through user control or real-world data. In the former, trajectory-based user control signals can be converted into guiding trajectories for simulation \cite{Shen_data_2018}. Predefined crowd motion `patches' can be used to compose heterogeneous crowd motions \cite{Jordao_crowd_2014}. The purpose of this kind of guidance is to give the user the full control to `sculpture' crowd motions. The latter is to guide simulations using real-world data to mimic real crowd motions. Given data and a parameterized simulation model, optimizations are used to fit the model on the data \cite{wolinski_parameter_2014}. Alternatively, features can be extracted and compared for different simulations, so that predictions can be made about different steering methods on a simulation task \cite{karamouzas_crowd_2019}. Our approach also heavily relies on data and is thus similar to the latter. But instead of anchoring on the modelling of individual motions, it focuses on the analysis of scene semantics/activities. It also considers intrinsic motion randomness in a structured and principled way. 

\section{Methodology overview}

The overview of our framework is in \figref{teaser}. Without loss of generality, we assume that the input is raw trajectories/tracklets which can be extracted from videos by existing trackers, where we can estimate the temporal and velocity information. Naively modelling the trajectories/tracklets, e.g. by simple descriptive statistics such as average speed, will average out useful information and cannot capture the data heterogeneity. To capture the heterogeneity in the presence of noise and randomness, we seek an underlying invariant as the scene descriptor. Based on empirical observations, steady space flows, characterized by groups of geometrically similar trajectories, can be observed in many crowd scenes. Each flow is a recurring activity connecting subspaces with designated functionalities, e.g. a flow from the front entrance to the ticket office then to a platform in a train station. Further, this flow reveals certain semantic information, i.e. people buying tickets before going to the platforms. Overall, all flows in a scene form a good basis to describe the crowd activities and the basis is an underlying invariant. How to compute this basis is therefore vital in analysis.

However, computing such a basis is challenging. Naive statistics of trajectories are not descriptive enough because the basis consists of many flows, and is therefore highly heterogeneous and multi-modal. Further the number of flows is not known \textit{a priori}. Since the flows are formed by groups of geometrically similar trajectories/tracklets, a natural solution is to cluster them \cite{Bian_2018_survey}. In this specific research context, unsupervised clustering is needed due to that the shear data volume prohibits human labelling. In unsupervised clustering, popular methods such as K-means and Gaussian Mixture Models \cite{bishop_pattern_2007} require a pre-defined cluster number which is hard to know in advance. Hierarchical Agglomerative Clustering \cite{Kaufman_2005_introduction} does not require a predefined cluster number, but the user must decide when to stop merging, which is similarly problematic. Spectral-based clustering methods \cite{Shi_2000_normalized} solve this problem, but require the computation of a similarity matrix whose space complexity is $O(n^2)$ on the number of trajectories. Too much memory is needed for large datasets and performance degrades quickly with increasing matrix size. Due to the afore-mentioned limitations, non-parametric Bayesian approaches were proposed \cite{wang_path_2016,wang_trending_2017}. However, a new approach is still needed because the previous approaches only consider space, and therefore cannot be reused or adapted for our purposes.

We propose a new non-parametric Bayesian method to cluster the trajectories with the time and velocity information in an \textit{unsupervised} fashion, which requires neither manual labelling nor the prior knowledge of cluster number. The outcome of clustering is a series of modes, each being a unique distribution over space, time and speed. Then we propose new methods for data visualization, simulation evaluation and automated simulation guidance.

We first introduce the background of one family of non-parametric Bayesian models, Dirichlet Processes (DPs), and Hierarchical Dirichlet Processes (HDP) (\secref{background}). We then introduce our new model Triplet HDPs (\secref{THDP}) and new inference method Chinese Restaurant Franchise League (\secref{inference}). Finally new methods are proposed for visualization (\secref{vis}), comparison (\secref{metrics}) and simulation guidance (\secref{simGuidance}).

\section{Our Method}
\subsection{Background}
\label{sec:background}
{\bf Dirichlet Process}. To understand DP, imagine there is a multi-modal 1D dataset with five high-density areas (modes). Then a classic five-component Gaussian Mixture Model (GMM) can fit the data via Expectation-Minimization \cite{bishop_pattern_2007}. Now further generalize the problem by assuming that there are an unknown number of high-density areas. In this case, an ideal solution would be to impose a prior distribution which can represent an infinite number of Gaussians, so that the number of Gaussians needed, their means and covariances can be automatically learnt. DP is such a prior. 

A DP($\gamma$, H) is a probabilistic measure on measures \cite{Ferguson_bayyesian_1973}, with a scaling parameter $\gamma$ > 0 and a base probability measure $H$. A draw from DP, $G$ \textasciitilde $DP(\gamma, H)$ is: $G = \sum_{k=1}^{\infty} \beta_k \delta_{\phi_k}$, where $\beta_k \in \bm{\beta}$ is random and dependent on $\gamma$. $\phi_k \in \bm{\phi}$ is a variable distributed according to $H$, $\phi_k \sim H$. $\delta_{\phi_k}$ is called an atom at $\phi_k$. Specifically for the example problem above, we can define $H$ to be a Normal-Inverse-Gamma (NIG) so that any draw, $\phi_k$, from $H$ is a Gaussian, then $G$ becomes an Infinite Gaussian Mixture Model (IGMM) \cite{Rasmussen_infinite_1999}. In practice, $k$ is finite and computed during inference.

\begin{figure}[t]
    \centering
    \includegraphics[width=1.\linewidth]{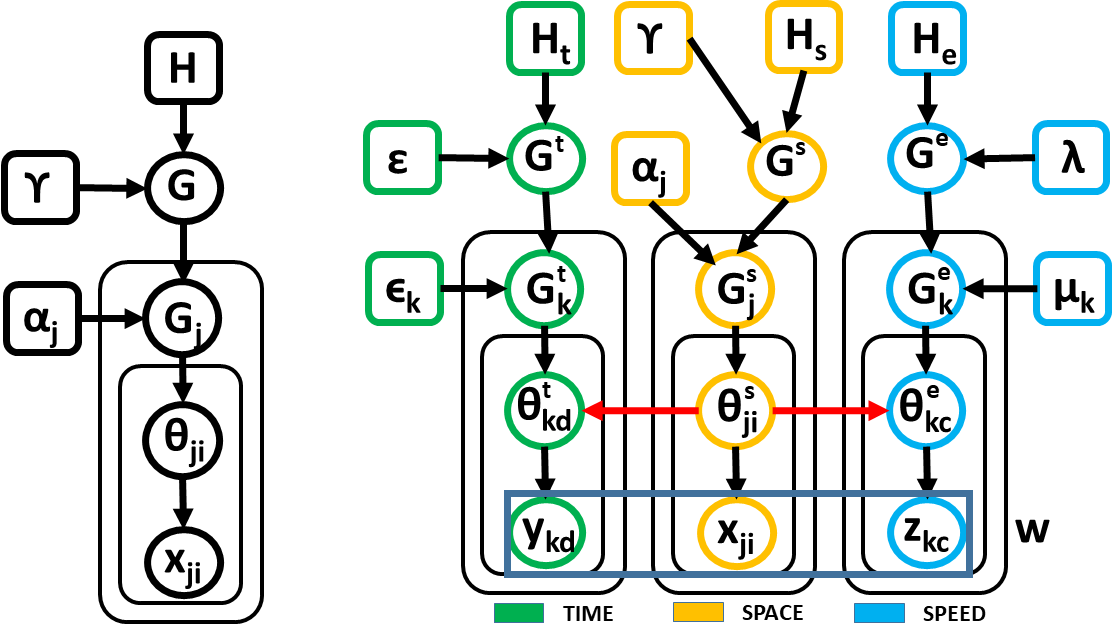}
    \caption{Left: HDP. Right: Triplet HDP.}
    \label{fig:models}
    \vspace{-1em}
\end{figure}

{\bf Hierarchical DPs}. Now imagine that the multi-modal dataset in the example problem is observed in separate data groups. Although all the modes can be observed from the whole dataset, only a subset of the modes can be observed in any particular data group. To model this phenomenon, a parent DP is used to capture all the modes with a child DP modelling the modes in each group:
\begin{equation}
\label{eq:HDP}
G_j \sim DP(\alpha_j, G) \; or \; G_j = \sum_{i=1}^{\infty} \beta_{ji} \delta_{\psi_{ji}} \ \ \text{where} \ \ \ G = \sum_{k=1}^{\infty} \beta_k \delta_{\phi_k}
\end{equation}
where $G_j$ is the modes in the $j$th data group. $\alpha_j$ is the scaling factor and $G$ is its based distribution. $\beta_{ji}$ is the weight and $\delta_{\psi_{ji}}$ is the atom. Now we have the Hierarchical DPs, or HDP \cite{teh_hierarchical_2006} (\figref{models} Left). At the top level, the modes are captured by $G \sim DP(\gamma, H)$. In each data group $j$, the modes are captured by $G_j$ which is dependent on $\alpha_j$ and $G$. This way, the modes, $G_j$, in every data group come from the common set of modes $G$, i.e. $\psi_{ji} \in \{\phi_1, \phi_2,...,\phi_k\}$. In \figref{models} Left, there is also a variable $\theta_{ji}$ called \textit{factor} which indicates with which mode ($\psi_{ji}$ or equally $\phi_k$) the data sample $x_{ji}$ is associated. Finally, if $H$ is again a NIG prior, then the HDP becomes Hierarchical Infinite Gaussian Mixture Model (HIGMM).

\subsection{Triplet-HDPs (THDP)}
\label{sec:THDP}
We now introduce THDP (\figref{models} Right). There are three HDPs in THDP, to model space, time and speed. We name them Time-HDP (Green), Space-HDP (Yellow) and Speed-HDP (Blue). Space-HDP is to compute space modes. Time-HDP and Speed-HDP are to compute the time and speed modes associated with each space mode, which requires the three HDPs to be linked. The modeling choice of the links will be explained later. The only observed variable in THDP is $w$, an observation of a person in a frame. It includes a location-orientation ($x_{ji}$), timestamp ($y_{kd}$) and speed ($z_{kc}$). $\theta^s_{ji}$, $\theta^t_{kd}$ and $\theta^e_{kc}$ are their factor variables. Given a single observation denoted as $w$, we denote one trajectory as $\bar{w}$, a group of trajectories as $\check{w}$ and the whole data set as $\bm{w}$. Our final goal is to compute the space, time and speed modes, given $\bm{w}$:

\begin{equation}
\label{eq:THDPTop}
G^s = \sum_{k=1}^{\infty} \beta_k \delta_{\phi^s_k}\ \ \ \ \ \ \ G^t = \sum_{l=1}^{\infty} \zeta_l \delta_{\phi^t_l}\ \ \ \ \ \ \ G^e = \sum_{q=1}^{\infty} \rho_q \delta_{\phi^e_q}
\end{equation}

In THDP, a space mode is defined to be a group of geometrically similar trajectories $\check{w}$. Since these trajectories form a flow, we also refer to it as a space flow. A space flow's timestamps ($y_{kd}$s) and speed ($z_{kc}$s) are both 1D data and can be modelled in similar ways. We first introduce the Time-HDP. One space flow $\check{w}$ might appear, crescendo, peak, wane and disappear several times. If a Gaussian distribution is used to represent one time peak on the timeline, multiple Gaussians are needed. Naturally IGMM is used to model the $y_{kd}\in\check{w}$. A possible alternative is to use Poisson Processes to model the entry time. But IGMM is chosen due to its ability to fit complex multi-modal distributions. It can also model a flow for the entire duration. Next, since there are many space flows and the $y_{kd}$s of each space flow form a timestamp data group, we therefore assume that there is a common set of time peaks shared by all space flows and each space flow shares only a subset. This way, we use a DP to represent all the time peaks and a child DP below the first DP to represent the peaks in each space flow. This is a HIGMM (for the Time-HDP) where the $H_t$ is a NIG. Similarly for the speed, $z_{kc}\in\check{w}$ can also have multiple peaks on the speed axis, so we use IGMM for this. Further, there are many space flows. We again assume that there is a common set of speed peaks and each space flow only has a subset of these peaks and use another HIGMM for the Speed-TDP.

\begin{figure}[tb]
    \centering
    \includegraphics[width=1.\linewidth]{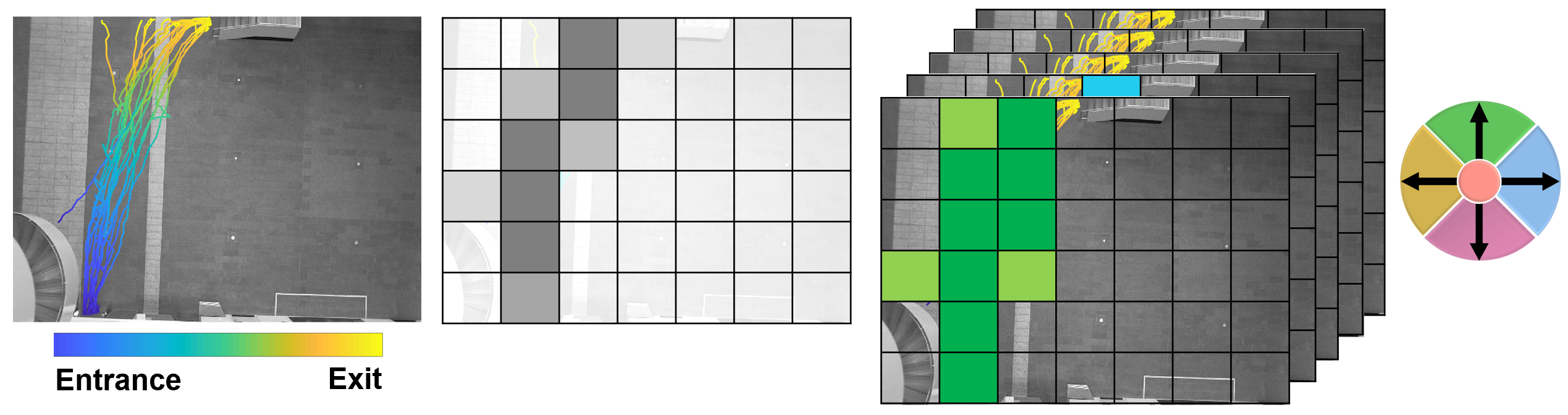}
    \caption{From left to right: 1. A space flow. 2. Discretization and flow cell occupancy, darker means more occupants. 3. Codebook with normalized occupancy as probabilities indicated by color intensities. 4. Five colored orientation subdomains (Pink indicates static).}
    \label{fig:discretization}
    \vspace{-1em}
\end{figure}

After Time-HDP and Speed-HDP, we introduce the Space-HDP. The Space-HDP is different because, unlike time and speed, space data ($x_{ji}$s) is 4D (2D location + 2D orientation), which means its modes are also multi-dimensional. In contrast to time and speed, a 4D Gaussian cannot represent a group of similar trajectories well. So we need to use a different distribution. Similar to \cite{wang_trending_2017}, we discretize the image domain (\figref{discretization}: 1) into a m $\times$ n grid (\figref{discretization}: 2). The discretization serves three purposes: 1. the cell occupancy serves as a good feature for a flow, since a space flow occupies a fixed group of cells. 2. it removes noises caused by frequent turns and tracking errors. 3. it eliminates the dependence on full trajectories. As long as instantaneous positions and velocities can be estimated, THDP can cluster observations. This is crucial in dealing with real-world data where full trajectories cannot be guaranteed. Next, since there is no orientation information so that the representation cannot distinguish between flows from A-to-B and flows from B-to-A, we discretize the instantaneous orientation into 5 cardinal subdomains (\figref{discretization}: 4). This makes the grid m $\times$ n $\times$ 5 (\figref{discretization}: 3), which now becomes a \textit{codebook} and every 4D $x_{ji}$ can be converted into a cell occupancy. Note although the grid resolution is problem-specific, it does not affect the validity of our method.

Next, since the cell occupancy on the grid (after normalization) can be seen as a Multinomial distribution, we use Multinomials to represent space flows. This way, a space flow has high probabilities in some cells and low probabilities in others (\figref{discretization}:3). Further, we assume the data is observed in groups and any group could contain multiple flows. We use a DP to model all the space flows of the whole dataset with child DPs representing the flows in individual data groups, e.g. video clips. This is a HDP (Space-HDP) with $H_s$ being a Dirichlet distribution.

After the three HDPs introduced separately, we need to link them, which is the key of THDP. For a space flow $\check{w_1}$, all $x_{ji}\in\check{w_1}$ are associated with the same space mode, denoted by $\phi^s_1$, and all $y_{kd}\in\check{w_1}$ are associated with the time modes \{$\phi^t_1$\} which forms a temporal profile of $\phi^s_1$. This indicates that $y_{kd}$'s time mode association is dependent on $x_{ji}$'s space mode association. In other words, if $x_{ji}^1\in\check{w}_1$ ($\phi^s_1$) and $x_{ji}^2\in\check{w}_2$ ($\phi^s_2$), where $x_{ji}^1 = x_{ji}^2$ but $\check{w}_1\ne\check{w}_2$ (two flows can partially overlap), then their corresponding $y_{kd}^1\in\check{w}_1$ and $y_{kd}^2\in\check{w}_2$ should be associated with \{$\phi^t_1$\} and \{$\phi^t_2$\} where \{$\phi^t_1$\} $\neq$ \{$\phi^t_2$\} when $\check{w}_1$ and $\check{w}_2$ have different temporal profiles. We therefore condition $\theta^t_{kd}$ on $\theta^s_{ji}$ (The left red arrow in \figref{models} Right) so that $y_{kd}$'s time mode association is dependent on $x_{ji}$'s space mode association. Similarly, a conditioning is also added to $\theta^e_{kc}$ on $\theta^s_{ji}$. This way, $w$'s associations to space, time and speed modes are linked. This is the biggest feature that distinguishes THDP from just a simple collection of HDPs, which would otherwise require doing analysis on space, time and dynamics separately, instead of holistically.

\section{Inference}
\label{sec:inference}
Given data $\bm{w}$, the goal is to compute the posterior distribution $p$(\bm{$\beta$}, \bm{$\phi^s$}, \bm{$\zeta$}, \bm{$\phi^t$}, \bm{$\rho$}, \bm{$\phi^e$} | {\bf w}). Existing inference methods for DPs include MCMC \cite{teh_hierarchical_2006}, variational inference \cite{hoffman_stochastic_2013} and geometric optimization \cite{Yurochkin_geometric_2016}. However, they are designed for simpler models (e.g. a single HDP). Further, both variational inference and geometric optimization suffer from local minimum. We therefore propose a new MCMC method for THDP. The method is a major generalization of Chinese Restaurant Franchise (CRF). Next, we first give the background of CRF, then introduce our method.

\subsection{Chinese Restaurant Franchise (CRF)}
\label{sec:CRF}

A single DP has a Chinese Restaurant Process (CRP) representation. CRF is its extension onto HDPs. We refer the readers to \cite{teh_hierarchical_2006} for details on CRP. Here we directly follow the CRF metaphor on HDP (\eqref{HDP}, \figref{models} Left) to compute the posterior distribution $p$(\bm{$\beta$}, \bm{$\phi$} | {\bf x}). In CRF, each observation $x_{ji}$ is called a \textit{customer}. Each data group is called a \textit{restaurant}. Finally, since a customer is associated with a mode (indicated by $\theta_{ji}$), the mode is called a \textit{dish} and is to be learned, as if the customer ordered this dish. CRF dictates that, in every restaurant, there is a potentially infinite number of tables, each with only one dish and many customers sharing that dish. There can be multiple tables serving the same dish. All dishes are on a global menu shared by all restaurants. The global menu can also contain an infinite number of dishes. In summary, we have multiple restaurants with many tables where customers order dishes from a common menu.

CRF is a Gibbs sampling approach. The sampling process is conducted at both customer and table level alternatively. At the customer level, each customer is treated, in turn, as a new customer, given all the other customers sitting at their tables. Then she needs to choose a table in her restaurant. There are two criteria influencing her decision: 1. how many customers are already at the table (\textit{table popularity}) and 2. how much she likes the dish on that table (\textit{dish preference}).  If she decides to not sit at any existing table, she can create a new table then order a dish. This dish can be from the menu or she can create a new dish and add it to the menu. Next, at the table-level, for each table, all the customers sitting at that table are treated as a new group of customers, and are asked to choose a dish together. Their collective dish preference and how frequently the dish is ordered in all restaurants (dish popularity) will influence their choice. They can choose a dish from the menu or create a new one and add it to the menu. We give the algorithm in \algref{CRF} and refer the readers to Appx. \ref{sec:AppCRF} for more details.

\setlength{\textfloatsep}{0pt}

\begin{algorithm}[tb]
\SetAlgoLined
\KwResult{\bm{$\beta$}, \bm{$\phi$} (\eqref{HDP})}
Input: \bm{$x$} \;
\While{Not converged}{
    \For{every restaurant j}{
        \For{every customer $x_{ji}$}{
            Sample a table $t_{ji}$ (\eqref{tableSampling}, Appx. \ref{sec:AppCRF})\;
            
            \If{a new table is chosen}{
                Sample a dish or create a new dish (\eqref{dishSampling}, Appx. \ref{sec:AppCRF})
            }
        }
        
        \For{every table and its customers \bm{$x_{jt}$}}{
            Sample a new dish (\eqref{tableDishSampling}, Appx. \ref{sec:AppCRF})
        }
    }
    
    Sample hyper-parameters \cite{teh_hierarchical_2006}
    
}
\caption{Chinese Restaurant Franchise}
\label{alg:CRF}
\end{algorithm}

\subsection{Chinese Restaurant Franchise League (CRFL)}
\label{sec:CRFL}
We generalize CRF by proposing a new method called Chinese Restaurant Franchise League. We first change the naming convention by adding prefixes space-, time- and speed- to customers, restaurant and dishes to distinguish between corresponding variables in the three HDPs. For instance, an observation $w$ now contains a space-customer $x_{ji}$, a time-customer $y_{kd}$ and a speed-customer $z_{kc}$. CRFL is a Gibbs sampling scheme, shown in \algref{CRFL}. The differences between CRF and CRFL are on two levels. At the top level, CRFL generalizes CRF by running CRF alternatively on three HDPs. This makes use of the conditional independence between the Time-HDP and the Speed-HDP given the Space-HDP fixed. At the bottom level, there are {\bf three} major differences in the sampling, between \eqref{tableSampling} and \eqref{CRFLTable}, \eqref{dishSampling} and \eqref{CRFLDishSampling}, \eqref{tableDishSampling} and \eqref{CRFLTableDishSampling}.

\begin{algorithm}
\SetAlgoLined
\KwResult{\bm{$\beta$}, \bm{$\phi^s$}, \bm{$\zeta$}, \bm{$\phi^t$}, \bm{$\rho$}, \bm{$\phi^e$} (\eqref{THDPTop})}
Input: \bm{$w$} \;
\While{Not converged}{
    Fix all variables in Space-HDP\;
    Do one CRF iteration (line 3-13, {\bf\algref{CRF}}) on Time-HDP\;
    Do one CRF iteration (line 3-13, {\bf\algref{CRF}}) on Speed-HDP\;
    \For{every space-restaurant j in Space-HDP}{
        \For{every space-customer $x_{ji}$}{
            Sample a table $t_{ji}$ (\eqref{CRFLTable})\;
            
            \If{a new table is chosen}{
                Sample a dish or create a new dish (\eqref{CRFLDishSampling})\;
            }
        }
        
        \For{every table and its space-customers \bm{$x_{jt}$}}{
            Sample a new space-dish (\eqref{CRFLTableDishSampling})\;
        }
    }
    Sample hyper-parameters (Appx. \ref{sec:APPHyperParameters})\;
}
\caption{Chinese Restaurant Franchise League}
\label{alg:CRFL}
\end{algorithm}

The first difference is when we do customer-level sampling (line 8 in \algref{CRFL}), the left side of \eqref{tableSampling} in CRF becomes:

\begin{equation}
\label{eq:CRFLTable}
p(t_{ji} = t, x_{ji}, y_{kd}, z_{kc} | \mathbf{x^{-ji}}, \mathbf{t^{-ji}}, \mathbf{k}, \mathbf{y^{-kd}}, \mathbf{o^{-kd}}, \mathbf{l}, \mathbf{z^{-kc}}, \mathbf{p^{-kc}}, \mathbf{q}) 
\end{equation}
where $t_{ji}$ is the new table for space-customer $x_{ji}$. $y_{kd}$ and $z_{kc}$ are the time and speed customer. $\mathbf{x^{-ji}}$ and $\mathbf{t^{-ji}}$ are the other customers (excluding $x_{ji}$) in the $j$th space-restaurant and their choices of tables. $\mathbf{k}$ is the space dishes. Correspondingly, $\mathbf{y^{-kd}}$ and $\mathbf{o^{-kd}}$ are the other time-customers (excluding $y_{kd}$) in the $k$th time-restaurant and their choices of tables. $\mathbf{l}$ is the time dishes. Similarly, $\mathbf{z^{-kc}}$ and $\mathbf{p^{-kc}}$ are the other speed-customers (excluding $z_{kc}$) in the $k$th speed-restaurant and their choices of tables. $\mathbf{q}$ is the speed-dishes. The intuitive interpretation of the differences between \eqref{CRFLTable} and \eqref{tableSampling} is: when a space-customer $x_{ji}$ chooses a table, the popularity and preference are not the only criteria anymore. She has to also consider the preferences of her associated time-customer $y_{kd}$ and speed-customer $z_{kc}$. This is because when $x_{ji}$ orders a different space-dish, $y_{kd}$ and $z_{kc}$ will be placed into a different time-restaurant and speed-restaurant, due to that the organizations of time- and speed-restaurants are dependent on the space-dishes (the dependence of $\theta^t_{kd}$ and $\theta^e_{kc}$ on $\theta^s_{ji}$). Each space-dish corresponds to a time-restaurant and a speed-restaurant (see \secref{THDP}). Since a space-customer's choice of space-dish can change during CRFL, the organization of time- and speed-restaurants becomes dynamic! This is why CRF cannot be directly applied to THDP. 

The second difference is when we need to sample a dish (line 10 in \algref{CRFL}), the left side of \eqref{dishSampling} in CRF becomes:
\ 
\begin{multline}
\label{eq:CRFLDishSampling}
p(k_{jt^{new}} = k, x_{ji}, y_{kd}, z_{kc} | \mathbf{k^{-jt^{new}}}, \mathbf{y^{-kd}}, \mathbf{o^{-kd}},\\ \mathbf{l}, \mathbf{z^{-kc}}, \mathbf{p^{-kc}}, \mathbf{q}) \propto \\
\begin{cases}
&m_{\cdot k}p(x_{ji}|\cdots)p(y_{kd}|\cdots)p(z_{kc}|\cdots) \\
&\gamma p(x_{ji}|\cdots)p(y_{kd}|\cdots)p(z_{kc}|\cdots)
\end{cases}
\vspace{-2em}
\end{multline}
where $k_{jt^{new}}$ is the new dish for customer $x_{ji}$. $\cdots$ represents all the conditional variables for simplicity. $p(y_{kd}|\cdots)$ and $p(z_{kc|\cdots})$ are the major differences. We refer the readers to Appx. \ref{sec:AppCRFL} regarding the computation of \eqref{CRFLTable} and \eqref{CRFLDishSampling}. 

The last difference is when we do the table-level sampling (line 14 in \algref{CRFL}), the left side of \eqref{tableDishSampling} in CRF changes to:
\begin{multline}
\label{eq:CRFLTableDishSampling}
p(k_{jt} = k, \mathbf{x_{jt}}, \mathbf{y_{kd_{jt}}}, \mathbf{z_{kc_{jt}}} | \mathbf{k^{-jt}}, \mathbf{y^{-kd_{jt}}}, \mathbf{o^{-kd_{jt}}},\\ \mathbf{l^{-ko}}, \mathbf{z^{-kc_{jt}}}, \mathbf{p^{-kc_{jt}}}, \mathbf{q^{-kp}}) \propto \\
\begin{cases}
&m_{\cdot k}^{-jt}p(\mathbf{x_{jt}}|\cdots)p(\mathbf{y_{kd_{jt}}}|\cdots)p(\mathbf{z_{kc_{jt}}}|\cdots) \\
&\gamma p(\mathbf{x_{jt}}|\cdots)p(\mathbf{y_{kd_{jt}}}|\cdots)p(\mathbf{z_{kc_{jt}}}|\cdots)
\end{cases}
\end{multline}
where $\mathbf{x_{jt}}$ is the space-customers at the $t$th table, $\mathbf{y_{kd_{jt}}}$ and $\mathbf{z_{kc_{jt}}}$ are the associated time- and speed-customers. $\mathbf{k^{-jt}}$, $\mathbf{y^{-kd_{jt}}}$, $\mathbf{o^{-kd_{jt}}}$, $\mathbf{l^{-ko}}$, $\mathbf{z^{-kc_{jt}}}$, $\mathbf{p^{-kc_{jt}}}$, $\mathbf{q^{-kp}}$ are the rest and their table and dish choices in three HDPs. $\cdots$ represents all the conditional variables for simplicity. $p(\mathbf{x_{jt}}|\cdots)$ is the Multinomial $f$ as in \eqref{tableDishSampling}. Unlike \eqref{CRFLDishSampling},  $p(\mathbf{y_{kd_{jt}}}|\cdots)$ and $p(\mathbf{z_{kc_{jt}}}|\cdots)$ cannot be easily computed and needs special treatment. We refer the readers to Appx. \ref{sec:AppCRFL} for details.

Now we have fully derived CRFL. Given a data set {\bf w}, we can compute the posterior distribution $p$(\bm{$\beta$}, \bm{$\phi^s$}, \bm{$\zeta$}, \bm{$\phi^t$}, \bm{$\rho$}, \bm{$\phi^e$} | {\bf w})
where \bm{$\beta$}, \bm{$\zeta$} and \bm{$\rho$} are the weights of the space, time and speed dishes, \bm{$\phi^s$}, \bm{$\phi^t$} and \bm{$\phi^e$} respectively. \bm{$\phi^s$} are Multinomials. \bm{$\phi^t$} and \bm{$\phi^e$} are Gaussians. As mentioned in \secref{CRF}, the number of \bm{$\phi^s$}, is automatically learnt, so we do not need to know the space dish number in advance. Neither do we need it for \bm{$\phi^t$} and \bm{$\phi^e$}. This makes THDP non-parametric. Further, since one $\phi^s$ could be associated with potentially an infinite number of $\phi^t$s and $\phi^e$s and vice versa, the many-to-many associations are also automatically learnt.

\subsection{Time Complexity of CRFL}
For each sampling iteration in \algref{CRFL}, the time complexities of sampling on time-HDP, speed-HDP and space-HDP are $O[W(N+L)+KNL]$, $O[W(A+Q)+KAQ]$ and $O[W(M+K)+2W(K+1)\eta+JMK]$ respectively, where $\eta=N+L+A+Q$. $W$ is the total observation number. $K$, $L$ and $Q$ are the dish numbers of space, time and speed. $J$ is the number of space-restaurants. $M$, $N$ and $A$ are the average table numbers in space-, time- and speed-restaurants respectively. Note that $K$ appears in all three time complexities because the number of space-dishes is also the number of time- and space-restaurants. 

The time complexity of CRFL is $O[W(N+L)+KNL] + O[W(A+Q)+KAQ] + O[W(M+K)+2W(K+1)\eta+JMK]$. This time complexity is not high in practice. $W$ can be large, depending on the dataset, over which a sampling could be used to reduce the observation number. In addition, $K$ is normally smaller than 50 even for highly complex datasets. $L$ and $Q$ are even smaller. $J$ is decided by the user and in the range of 10-30. $M$, $N$ and $A$ are not large either due to the high aggregation property of DPs, i.e. each table tends to be chosen by many customers, so the table number is low.

\section{Visualization, Metrics and Simulation Guidance based on THDP}
THDP provides a powerful and versatile base for new tools. In this section, we present three tools for structured visualization, quantitative comparison and simulation guidance.

\subsection{Flexible and Structured Crowd Data Visualization}
\label{sec:vis}
After inference, the highly rich but originally mixed and unstructured data is now structured. This is vital for visualization. It is immediately easy to visualize the time and speed modes as they are mixtures of univariate Gaussians. The space modes require further treatments because they are m$\times$n$\times$5 Multinomials and hard to visualize. We therefore propose to use them as classifiers to classify trajectories. After classification, we select representative trajectories for a clear and intuitive visualization of flows. Given a trajectory $\bar{w}$, we compute a \textit{softmax} function:
\begin{equation}
p_k(\bar{w}) = \frac{e^{p_k(\bar{w})}}{\sum_{k=1}^K e^{p_k(\bar{w})}}\ \text{$k\in$[1, K]}
\end{equation}
where $p_k(\bar{w})$ = $p(\bar{w} | \beta_k, \phi^s_k, \bm{\zeta_k}, \bm{\phi^t}, \bm{\rho_k}, \bm{\phi^e})$. $\phi^s_k$ and $\beta_k$ are the $k$th space mode and its weight. The others are the associated time and speed modes. The time and speed modes ($\bm{\phi^t}$ and $\bm{\phi^e}$) are associated with space flow $\phi^s_k$, with weights, $\bm{\zeta_k}$ and $\bm{\rho_k}$. $K$ is the total number of space flows. This way, we classify every trajectory into a space flow. Then we can visualize representative trajectories with high probabilities, or show anomaly trajectories with low probabilities.

In addition, since THDP captures all space, time and dynamics, there is a variety of visualization. A period of time can be represented by a weighted combination of time modes \{$\phi^t$\}. Assuming that the user wants to see what space flows are prominent during this period, we can visualize trajectories based on $\int_{\bm{\rho},\bm{\phi^e}} p(\bm{\beta}, \bm{\phi^s} | \{\phi^t\})$, which gives the space flows with weights. This is very useful if for instance \{$\phi^t$\} is rush hours, $\int_{\bm{\rho},\bm{\phi^e}} p(\bm{\beta}, \bm{\phi^s} | \{\phi^t\})$ shows us what flows are prominent and their relative importance during the rush hours. Similarly, if we visualize data based on $\int_{\bm{\zeta},\bm{\phi^t}} p(\bm{\rho}, \bm{\phi^e}|\phi^s)$, it will tell us if people walk fast/slowly on the space flow $\phi^s$. A more complex visualization is $p(\bm{\zeta}, \bm{\phi^t},\bm{\rho},\bm{\phi^e} |\phi^s)$ where the time-speed distribution is given for a space flow $\phi^s$. This gives the speed change against time of this space flow, which could reveal congestion at times. 

Through marginalizing and conditioning on different variables (as above), there are many possible ways of visualizing crowd data and each of them reveals a certain aspect of the data. We do not enumerate all the possibilities for simplicity but it is very obvious that THDP can provide highly flexible and insightful visualizations.

\subsection{New Quantitative Evaluation Metrics}
\label{sec:metrics}
Being able to quantitatively compare simulated and real crowds is vital in evaluating the quality of crowd simulation. Trajectory-based \cite{guy_statistical_2012} and flow-based \cite{wang_path_2016} methods have been proposed. The first flow-based metrics are proposed in \cite{wang_path_2016} which is similar to our approach. In their work, the two metrics proposed were: average likelihood (AL) and distribution-pair distance (DPD) based on Kullback-Leibler (KL) divergence. The underlying idea is that a good simulation does not have to strictly reproduce the data but should have statistical similarities with the data. However, they only considered space. We show that THDP is a major generalization of their work and provides much more flexibility with a set of new AL and DPD metrics.

\subsubsection{AL Metrics}

Given a simulation data set, $\bm{\hat w} = (\bm{\hat x_{ji}}, \bm{\hat y_{kd}}, \bm{\hat z_{kc}})$
and $p$(\bm{$\beta$}, \bm{$\phi^s$}, \bm{$\zeta$}, \bm{$\phi^t$}, \bm{$\rho$}, \bm{$\phi^e$} | {\bf w}) inferred from real-world data $\bm{w}$, we can compute the AL metric based on space only, essentially computing the average space likelihood while marginalizing time and speed:

\begin{equation}
\label{eq:ALMetric}
\frac{1}{|\bm{\hat w}|}\sum_{j,i}\sum_{k=1}^{K} \beta_k\int_z\int_y p(\hat x_{ji} | \phi^s_k, \hat y_{kd}, \hat z_{kc})\ p(\hat y_{kd})p(\hat z_{kc})dydz
\end{equation}
where $|\bm{\hat w}|$ is the number of observations in $\bm{\hat w}$.  The dependence on \bm{$\beta$}, \bm{$\phi^s$}, \bm{$\zeta$}, \bm{$\phi^t$}, \bm{$\rho$}, \bm{$\phi^e$} are omitted for simplicity. If we completely discard time and speed, \eqref{ALMetric} changes to the AL metric in \cite{wang_trending_2017}, $\frac{1}{|\bm{\hat w}|}\sum_{j, i}\sum_{k}\beta_k p(\hat x_{ji} | \phi^s_k)$. However, the metric is just a special case of THDP. We give a list of AL metrics in \tabref{ALMetrics}, which all have similar forms as \eqref{ALMetric}.

\begin{table}[tb]
    \centering
    \begin{tabular}{l|c}
         Metric & To compare  \\\hline
        1.$\frac{1}{|\bm{\hat w}|}\sum p(\hat x_{ji}, \hat y_{kd}, \hat z_{kc} | \bullet)$ & overall similarity \\\hline
        2.$\frac{1}{|\bm{\hat w}|}\sum  p(\hat x_{ji}, \hat y_{kd}|\bullet)$  & space\&time ignoring speed \\\hline
        3.$\frac{1}{|\bm{\hat w}|}\sum  p(\hat x_{ji}, \hat z_{kc}|\bullet)$  & space\&speed ignoring time \\\hline
        4.$\frac{1}{|\bm{\hat w}|}\sum  p(\hat y_{kd}, \hat z_{kc}|\bullet)$  & time\&speed ignoring space \\\hline
        5.$\frac{1}{|\bm{\hat w}|}\sum p(\hat x_{ij}|\bullet)$  & space ignoring time \& speed \\\hline
        6.$\frac{1}{|\bm{\hat w}|}\sum p(\hat y_{kd}|\bullet)$  & time ignoring space \& speed \\\hline
        7.$\frac{1}{|\bm{\hat w}|}\sum\ p(\hat z_{kc}|\bullet)$  & speed ignoring space \& time \\\hline
    \end{tabular}
    \caption{AL Metrics, $\bullet$ represents \{$\bm{\beta}, \bm{\phi^s}, \bm{\zeta}, \bm{\phi^t}, \bm{\rho}, \bm{\phi^e}$\}.}
    \label{tab:ALMetrics}
\end{table}

\subsubsection{DPD Metrics}

AL metrics are based on average likelihoods, summarizing the differences between two data sets into one number. To give more flexibility, we also propose distribution-pair metrics. We first learn two posterior distributions $p$(\bm{$\hat \beta$}, \bm{$\hat \phi^s$}, \bm{$\hat \zeta$}, \bm{$\hat \phi^t$}, \bm{$\hat \rho$}, \bm{$\hat \phi^e$} | {\bm{$\hat w$}}) and $p$(\bm{$\beta$}, \bm{$\phi^s$}, \bm{$\zeta$}, \bm{$\phi^t$}, \bm{$\rho$}, \bm{$\phi^e$} | {\bf w}). Then we can compare individual pairs of $\phi^s$ and $\hat \phi^s$, $\phi^t$ and $\hat \phi^t$, $\phi^e$ and $\hat \phi^e$. Since all space, time and speed modes are probability distributions, we propose to use Jensen-Shannon divergence, as oppose to KL divergence \cite{wang_trending_2017} due to KL's asymmetry:

\begin{equation}
JSD(P || Q) = \frac{1}{2}D(P || M) + \frac{1}{2}D(Q || M)
\end{equation}
where $D$ is KL divergence and $M = \frac{1}{2}(P + Q)$. $P$ and $Q$ are probability distributions. Again, in the DPD comparison, THDP provides many options, similar to the AL metrics in \tabref{ALMetrics}. We only give several examples here. Given two space flows, $\phi^s$ and $\hat \phi^s$, JSD($\phi^s$ || $\hat \phi^s$) directly compares two space flows. Further, $P$ and $Q$ can be conditional distributions. If we compute JSD($p(\bm{\phi^t}$ | $\phi^s$) || p($\bm{\hat \phi^t}$ | $\hat \phi^s$)) where $\bm{\phi^t}$ and $\bm{\hat \phi^t}$ are the associated time modes of $\phi^s$ and $\hat \phi^s$ respectively. This is to compare the two temporal profiles. This is very useful when $\phi^s$ and $\hat \phi^s$ are two spatially similar flows but we want to compare the temporal similarity. Similarly, we can also compare their speed profiles JSD($p(\bm{\phi^e}$ | $\phi^s$) || p($\bm{\hat \phi^e}$ | $\hat \phi^s$)) or their time-speed profiles JSD($p(\bm{\phi^t}$, $\bm{\phi^e}$ | $\phi^s$) || p($\bm{\hat \phi^t}$, $\bm{\hat \phi^e}$ | $\hat \phi^s$)). In summary, similar to AL metrics, different conditioning and marginalization choices result in different DPD metrics.

\subsection{Simulation Guidance} 
\label{sec:simGuidance}
We propose a new method to automate simulation guidance with real-world data, which works with existing simulators including steering and global planning methods. Assuming that we want to simulate crowds in a given environment based on data, there are still several key parameters which need to be estimated including, starting/destination positions, the entry timing and the desired speed. After inferring, we use GMM to model both starting and destination regions for every space flow. This way, we completely eliminate the need for manual labelling, which is difficult in spaces with no designated entrances/exits (e.g. a square). Also, we removed the one-to-one mapping requirement of the agents in simulation and data. We can sample any number of agents based on space flow weights ($\bm{\beta}$) and still keep similar agent proportions on different flows to the data.  In addition, since each flow comes with a temporal and speed profile, we sample the entry timing and desired speed for each agent, to mimic the randomness in these parameters. It is difficult to manually set the timing when the duration is long and sampling the speed is necessary to capture the speed variety within a flow caused by latent factors such as different physical conditions.

Next, even with the right setting of all the afore-mentioned parameters, existing simulators tend to simulate straight lines whenever possible while the real data shows otherwise. This is due to that no intrinsic motion randomness is introduced. Intrinsic motion randomness can be observed in that people rarely walk in straight lines and they generate slightly different trajectories even when asked to walk several times between the same starting position and destination \cite{wang_trending_2017}. This is related to the state of the person as well as external factors such as collision avoidance. Individual motion randomness can be modelled by assuming the randomness is Gaussian-distributed \cite{guy_statistical_2012}. Here, we do not assume that all people have the same distribution. Instead, we propose to do a structured modelling. We observe that people on different space flows show different dynamics but share similar dynamics within the same flow. This is because people on the same flow share the same starting/destination regions and walk through the same part of the environment. In other words, they started in similar positions, had similar goals and made similar navigation decisions. Although individual motion randomness still exists, their randomness is likely to be similarly distributed. However, this is not necessarily true across different flows. We therefore assume that each space flow can be seen as generated by a unique dynamic system which captures the within-group motion randomness which implicitly considers factors such as collision avoidance. Given a trajectory, $\bar{w}$, from a flow $\check{w}$, we assume that there is an underlying dynamic system:

\begin{align}
\label{eq:LDS}
    x_t^{\bar{w}} = As_t + \omega_t \ \ \ \ \omega \sim N(0, \Omega) \nonumber \\
    s_t = Bs_{t-1} + \lambda_t \ \ \ \ \lambda \sim N(0, \Lambda)
\end{align}
where $x_t^{\bar{w}}$ is the observed location of a person at time $t$ on trajectory $\bar{w}$. $s_{t}$ is the latent state of the dynamic system at time $t$. $\omega_t$ and $\lambda_t$ are the observational and dynamics randomness. Both are white Gaussian noises. $A$ and $B$ are transition matrices. We assume that $\Omega$ is a known diagonal covariance matrix because it is intrinsic to the device (e.g. a camera) and can be trivially estimated. We also assume that $A$ is an identity matrix so that there is no systematic bias and the observation is only subject to the state $s_t$ and noise $\omega_t$. The dynamic system then becomes: $x_t^{\bar{w}} \sim N(Is_t, \Omega)$ and $s_t \sim N(Bs_{t-1}, \Lambda)$, where we need to estimate $s_t$, $B$ and $\Lambda$. Given the $U$ trajectories in $\check{w}$, the total likelihood is:

\begin{align}
    &p(\check{w}) = \Pi_{i=1}^U p(\bar{w}_i) \ \ \ \text{where}\nonumber \\
    &p(\bar{w}_i) = \Pi_{t=2}^{T_i-1} p(x^i_t|s_t) P(s_t|s_{t-1})\ \ \, s_1 = x^i_1, s_{T} = x^i_{T_i}
\end{align}
where $T_i$ is the length of trajectory $\bar{w}_i$. We maximize $log\ P(\check{w})$ via Expectation-Maximization \cite{bishop_pattern_2007}. Details can be found in the Appx. \ref{sec:AppSimGuidance}. After learning the dynamic system for a space flow and given a starting and destination location, $s_1$ and $s_T$, we can sample diversified trajectories while obeying the flow dynamics. During simulation guidance, one target trajectory is sampled for each agent and this trajectory reflects the motion randomness.

\begin{figure*}[tb]
\begin{subfigure}{\linewidth}
    \centering
    \includegraphics[width=\linewidth]{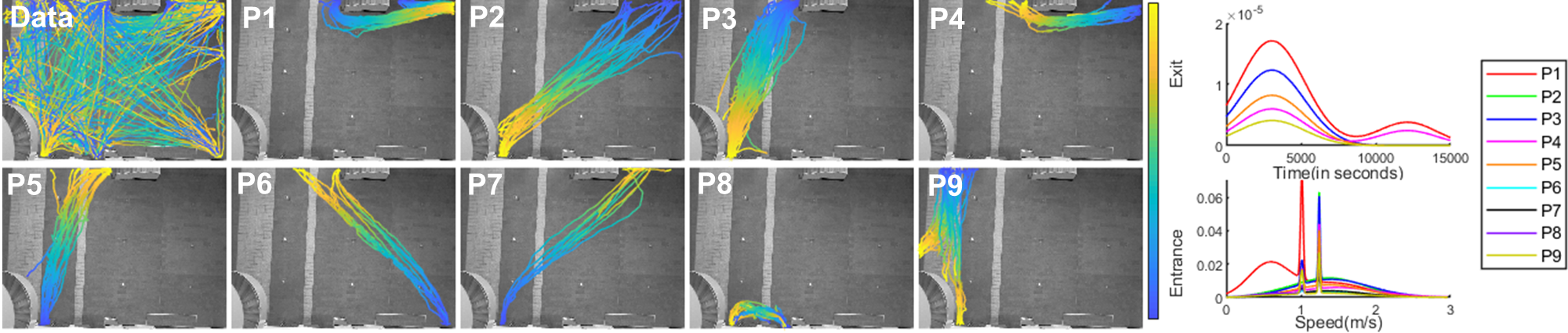}
\end{subfigure}\par\medskip
\begin{subfigure}{\linewidth}
    \centering
    \includegraphics[width=\linewidth]{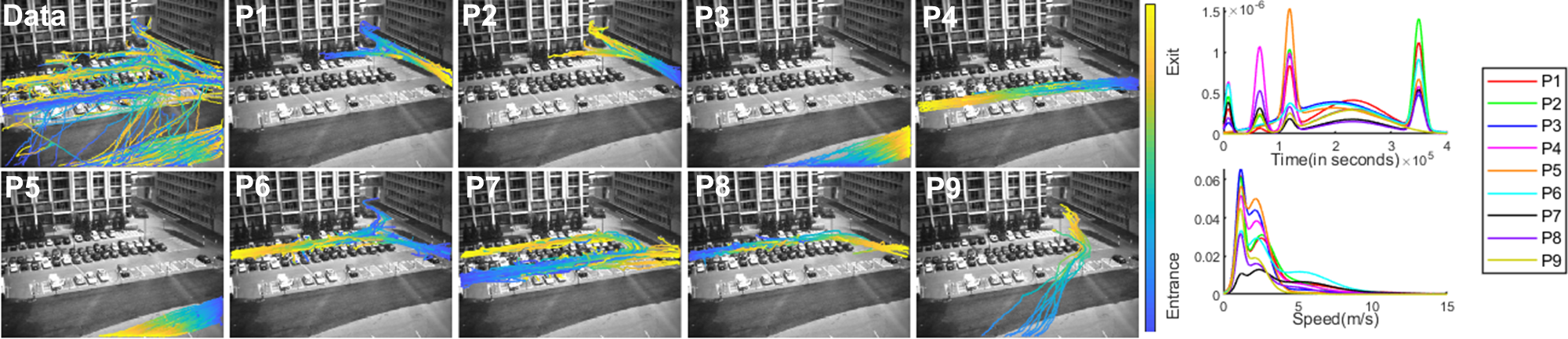}
\end{subfigure}\par\medskip
\begin{subfigure}{\linewidth}
    \centering
    \includegraphics[width=\linewidth]{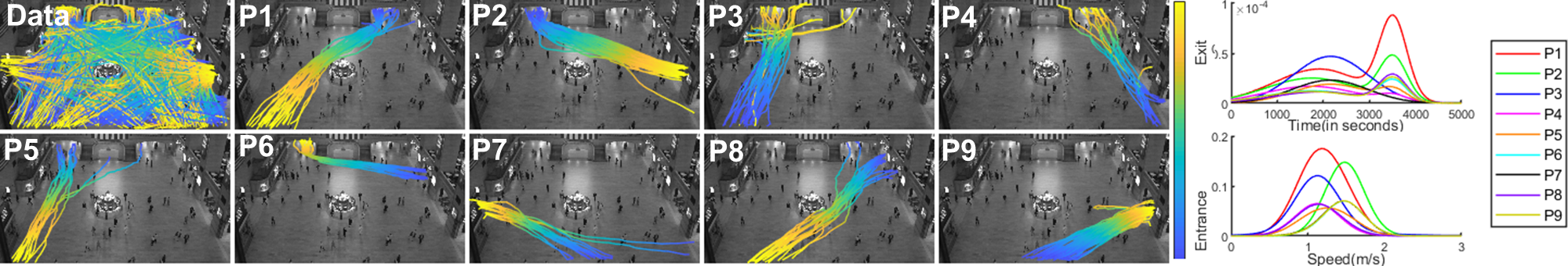}
\end{subfigure}
\caption{Forum (top), CarPark (Middle) and TrainStation (Bottom) dataset. In each dataset, Top left: original data; P1-P9: the top 9 space modes; Top right: the time modes of P1-P9; Bottom right: the speed modes of P1-P9. Both time and speed profiles are scaled by their respective space model weights, with the y axis indicating the likelihood.}
\label{fig:std_vis}
\end{figure*}

\section{Experiments}
In this section, we first introduce the datasets,  then show our highly informative and flexible visualization tool. Next, we give quantitative comparison results between simulated and real crowds by the newly proposed metrics. Finally, we show that our automated simulation guidance with high semantic fidelity. We only show representative results in the paper and refer the readers to the supplementary video and materials for details.

\subsection{Datasets}
We choose three publicly available datasets: {\bf Forum} \cite{majecka_statistical_2009}, {\bf CarPark} \cite{wang_trajectory_2008} and {\bf TrainStation} \cite{Yi_understanding_2015}, to cover different data volumes, durations, environments and crowd dynamics. Forum is an indoor environment in a school building, recorded by a top-down camera, containing 664 trajectories and lasting for 4.68 hours. Only people are tracked and they are mostly slow and casual. CarPark consists of videos of an outdoor car park with mixed pedestrians and cars, by a far-distance camera and contains totally 40,453 trajectories over five days. TrainStation is a big indoor environment with pedestrians and designated sub-spaces. It is from New York Central Terminal and contains totally 120,000 frames with 12,684 pedestrians within approximately 45 minutes. The speed varies among pedestrians. 

\subsection{Visualization Results}

We first show a general, full-mode visualization in \figref{std_vis}. Due to the space limit, we only show the top 9 space modes and their corresponding time and speed profiles. Overall, THDP is effective in decomposing highly mixed and unstructured data into structured results across different data sets. The top 9 space modes (with time and speed) are the main activities. With the environment information (e.g. where the doors/lifts/rooms are), the semantic meanings of the activities can be inferred. In addition, the time and dynamics are captured well. One peak of a space flow (indicated by color) in the time profiles indicates that this flow is likely to appear around that time. Correspondingly, one peak of a space flow in the speed profile indicates a major speed preference of the people on that flow. Multiple space flows can peak near one point in both the time and speed profiles. The speed profiles of Forum and TrainStation are slightly different, with most of the former distributed in a smaller region. This is understandable because people in TrainStation in general walk faster. The speed profile of CarPark is quite different in that it ranges more widely, up to 10m/s. This is because both pedestrians and vehicles were recorded.

\begin{figure}[t!]
\includegraphics[width=0.48\linewidth]{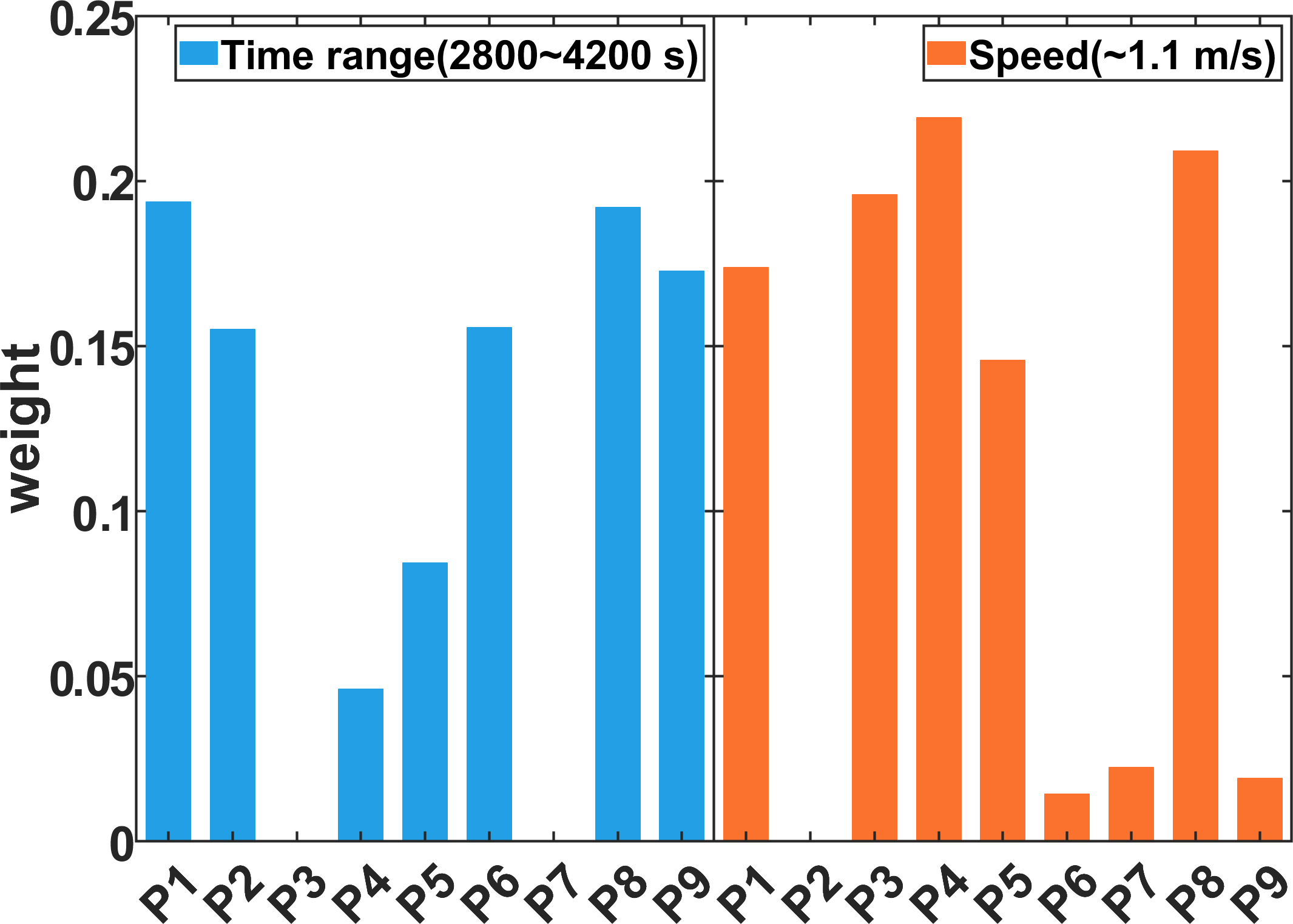}
\includegraphics[width=0.48\linewidth]{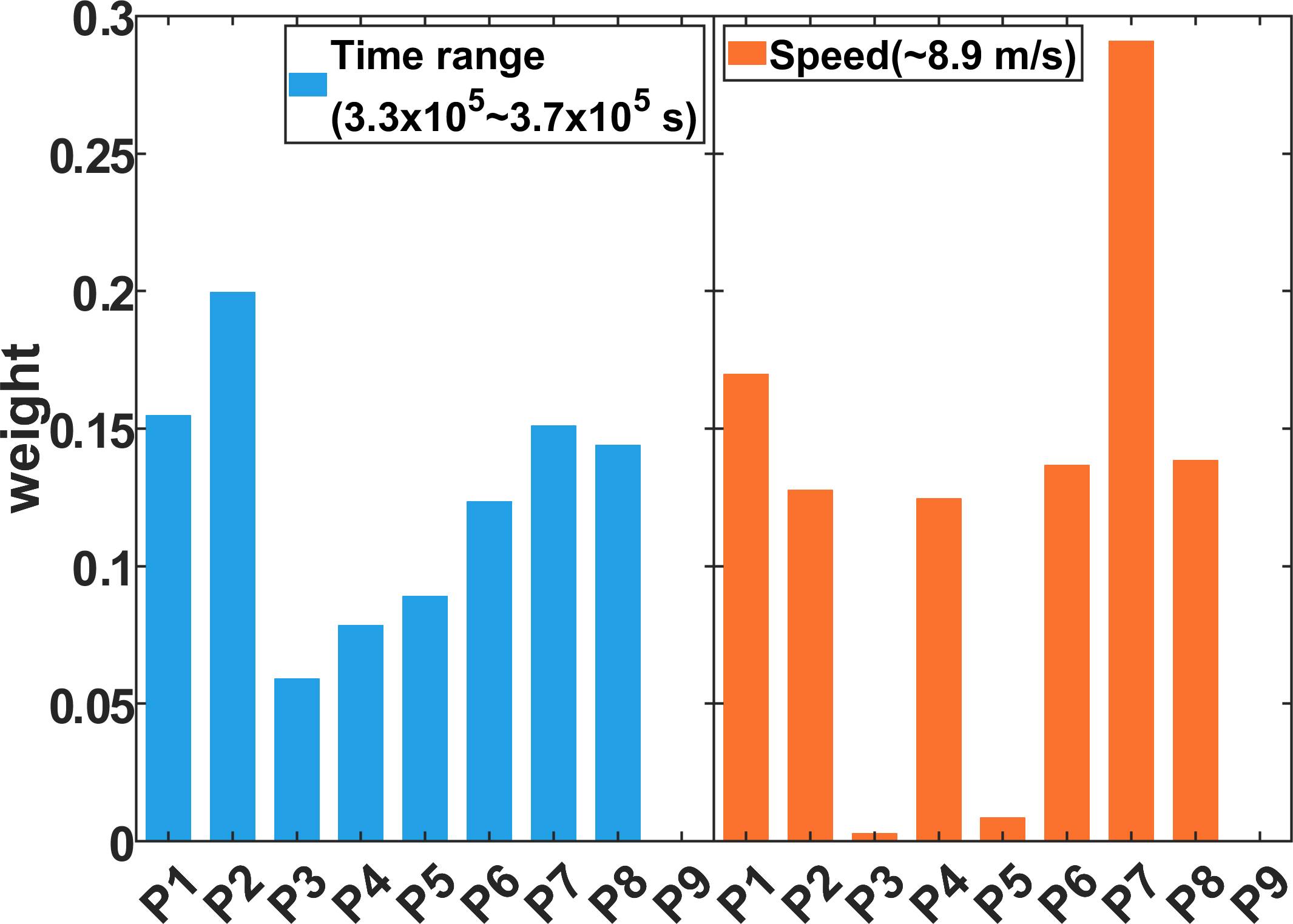}
\caption{Left: TrainStation, Right: CarPark. The space flow prominence (indicated by bar heights) of P1-P9 in \figref{std_vis} respectively given a time period (blue bars) or speed range (orange bars). The higher the bar, the more prominent the space flow is.}
\label{fig:time-speed_conditioned_vis}
\end{figure}

\begin{figure}[t!]
\centering
\includegraphics[width=1.\linewidth]{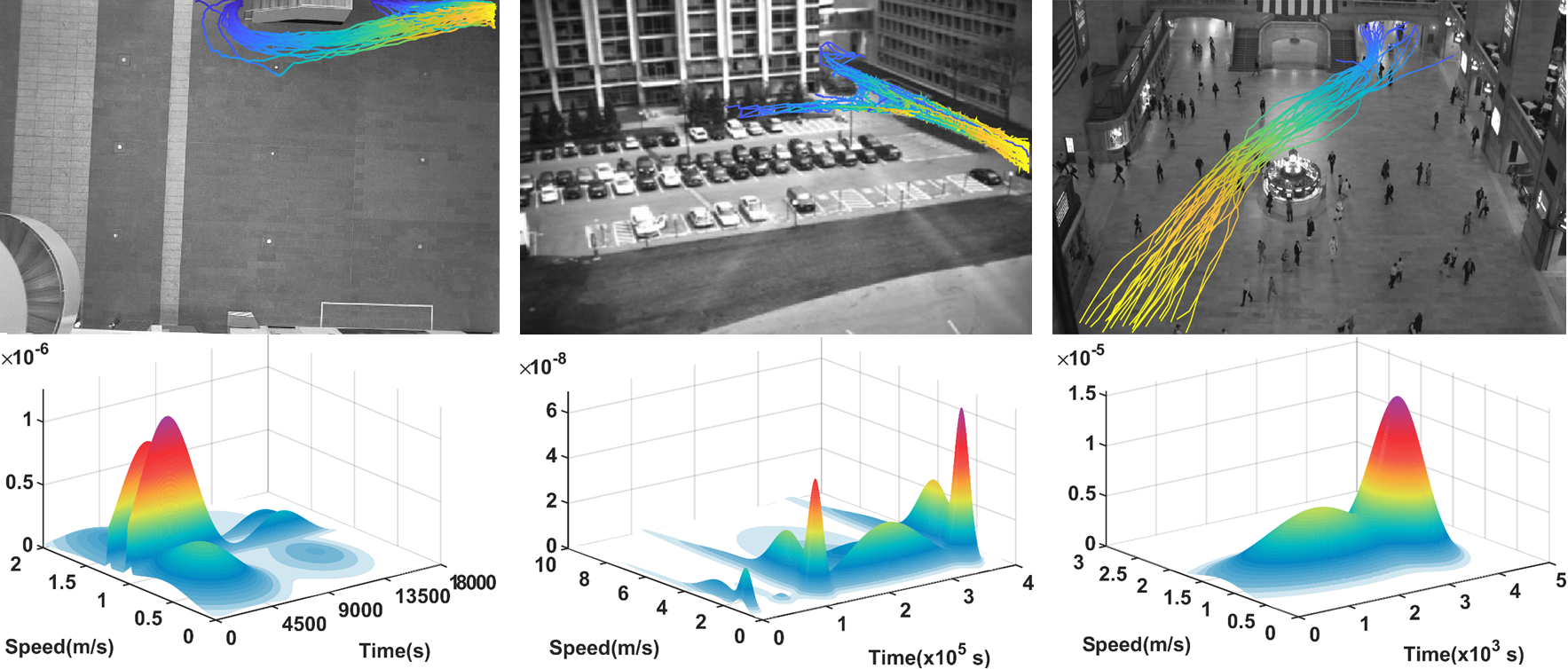}
\caption{Space flows from Forum, CarPark and TrainStation and their time-speed distributions. The y (up) axis is likelihood. The x and z axes are time and speed. The redder, the higher the likelihood is.}
\label{fig:space_conditioned_vis}
\end{figure}

Besides, we show conditioned visualization. Suppose that the user is interested in a period (e.g. rush hours) or speed range (e.g. to see where people generally walk fast/slowly), the associated flow weights can be visualized (\figref{time-speed_conditioned_vis}). This allows users to see which space flows are prominent in the chosen period or speed range. Conversely, given a space flow in interest, we can visualize the time-speed distribution (\figref{space_conditioned_vis}), showing how the speed changes along time, which could help identify congestion on that flow at times. 

\begin{figure}
\centering
\includegraphics[width=\linewidth]{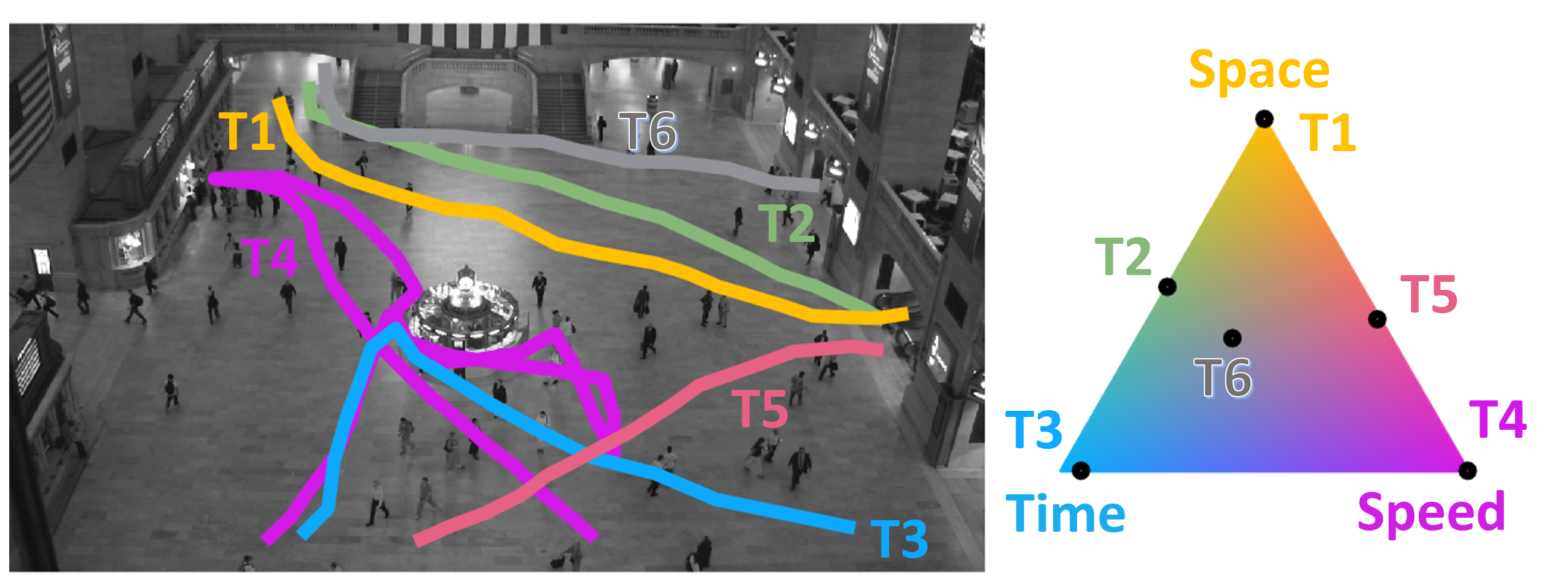}
\caption{Representative anomaly trajectories. Every trajectory has a corresponding location in the triangle on the right, indicating what factors contribute more in its abnormality. For instance, T1 is close to the space vertex, it means its spatial probability is relatively high and the main abnormality contribution comes from its time and speed. For T2, the contribution mainly comes from its speed.}
\label{fig:anomaly}
\end{figure}

Last but not least, we can identify anomaly trajectories and show unusual activities. The anomalies here refer to statistical anomalies. Although they are not necessarily suspicious behaviors or events, they can help the user to quickly reduce the number of cases needed to be investigated. Note that the anomaly is not only the spacial anomaly. It is possible that a spatially normal trajectory that is abnormal in time and/or speed. To distinguish between them, we first compute the probabilities of all trajectories and select anomalies. Then for each anomaly trajectory, we compute its relative probabilities (its probability divided by the maximal trajectory probability) in space, time and speed, resulting in three probabilities in [0, 1]. Then we use them (after normalization) as the bary-centric coordinates of a point inside of a colored triangle. This way, we can visualize what contributes to their abnormality (\figref{anomaly}). Take T1 for example. It has a normal spacial pattern, and therefore is close to the `space' vertex. It is far away from both `time' and `speed' vertex, indicating T1's time and speed patterns are very different from the others'. THDP can be used as a versatile and discriminative anomaly detector.

Non-parametric Bayesian approaches have been used for crowd analysis \cite{wang_path_2016,wang_trending_2017}. However, existing methods can be seen as variants of the Space-HDP and cannot decompose information in time and dynamics. Consequently, they cannot show any results related to time \& speed, as opposed to Fig. \ref{fig:std_vis}-\ref{fig:anomaly}. A naive alternative would be to use the methods in \cite{wang_path_2016,wang_trending_2017} to first cluster data regardless time and dynamics, then do per-cluster time and dynamics analysis, equivalent to using the Space-HDP first, then the time-HDP \& Speed-HDP subsequently. However, this kind of sequential analysis has failed due to one limitation: the spatial-only HDP misclassifies observations in the overlapped areas of flows \cite{Wang_Globally_2016}. The following time and dynamics analysis would be based on wrong clustering.  The simultaneity of considering all three types of information, accomplished by the links (red arrows in \figref{models} Right) among three HDPs in THDP, is therefore essential.

\subsection{Compare Real and Simulated Crowds}
\label{sec:comparison}
To compare simulated and real crowds, we ask participants (Master and PhD students whose expertise is in crowd analysis and simulation) to simulate crowds in Forum and TrainStation. We left CarPark out because its excessively long duration makes it extremely difficult for participants to observe. We built a simple UI for setting up simulation parameters including starting/destination locations, the entry timing and the desired speed for every agent. For simulator, our approach is agnostic about simulation methods. We chose ORCA in Menge \cite{Curtis_menge_2015} for our experiments but other simulation methods would work equally well. Initially, we provide the participants with only videos and ask them to do their best to replicate the crowd motions. They found it difficult because they had to watch the videos and tried to remember a lot of information, which is also a real-world problem of simulation engineers. This suggests that different levels of detail of the information are needed to set up simulations. The information includes variables such as entry timings and start/end positions, which are readily available, or descriptive statistics such as average speed, which can be relatively easily computed. We systematically investigate their roles in producing scene semantics. After several trials, we identified a set of key parameters including starting/ending positions, entry timing and desired speed. Different simulation methods require different parameters, but these are the key parameters shared by all. We also identified four typical settings where we gradually provide more and more information about these parameters. This design helps us to identify the qualitative and quantitative importance of the key parameters for the purpose of reproducing the scene semantics.

The first setting, denoted as Random, is where only the starting/destination regions are given. The participants have to estimate the rest. Based on Random, we further give the exact starting/ending positions, denoted by SDR. Next, we also give the entry timing for each agent based on SDR, denoted by SDRT. Finally, we give the average speed of each agent based on SDRT, denoted by SDRTS. Random is the least-informed scenario where the users have to estimate many parameters, while SDRTS is the most-informed situation. A comparison between the four settings is shown in \tabref{simSettings}.

\begin{table}[tb]
    \centering
    \begin{tabular}{c|c|c|c|c}
        Information / Setting & Random & SDR & SDRT & SDRTS \\
        \hline
        \small{Starting/Dest. Areas} & $\checkmark$ & $\checkmark$ & $\checkmark$ & $\checkmark$ \\
        \hline
        \small{Exact Starting/Dest. Positions} & $\times$ & $\checkmark$ & $\checkmark$ & $\checkmark$  \\
        \hline
        \small{Trajectory Entry Timing} & $\times$ & $\times$ & $\checkmark$ & $\checkmark$ \\
        \hline
        \small{Trajectory Average Speed} & $\times$ & $\times$ & $\times$ & $\checkmark$\\
        \hline
    \end{tabular}
    \caption{Different simulation settings and the information provided.}
    \label{tab:simSettings}
    \vspace{-1em}
\end{table}

We use four AL metrics to compare simulations with data, as they provide detailed and insightful comparisons: Overall (\tabref{ALMetrics}: 1), Space-Only (\tabref{ALMetrics}: 5), Space-Time (\tabref{ALMetrics}: 2) and Space-Speed (\tabref{ALMetrics}: 3) and show the comparisons in \tabref{AL_comparison}. In Random, the users had to guess the exact entrance/exit locations, entry timing and speed. It is very difficult to do by just watching videos and thus has the lowest score across the board. When provided with exact entrance/exit locations (SDR), the score is boosted in Overall and Space-Only. But the scores in Space-Time and Space-Speed remain relatively low. As more information is provided (SDRT \& SDRTS), the scores generally increase. This shows that our metrics are sensitive to space, time and dynamics information during comparisons. Further, each type of information is isolated out in the comparison. The Space-Only scores are roughly the same between SDR, SDRT and SDRTS. The Space-Time scores do not change much between SDRT and SDRTS. The isolation in comparisons makes our AL metrics ideal for evaluating simulations in different aspects, providing great flexibility which is necessary in practice.

\begin{table}[tb]
    \centering
    \begin{tabular}{l|c|c|c|c|c}
        Metric/Simulations &  Random & SDR & SDRT & SDRTS & Ours\\
        \hline
        Overall ($\times10^{-8}$) & 7.11 & 20.67 & 37.08 & 40.55& {\bf 57.9}  \\
        \hline
        Space-Only ($\times10^{-3}$) & 2.7 & 5.3 & 5.3 & 5.5 & 5.1\\ 
        \hline
        Space-Time ($\times10^{-7}$) & 1.23 & 2.96 & 5.56 & 5.77 & {\bf 6.02}\\ 
        \hline
        Space-Speed ($\times10^{-3}$) & 1.5 & 3.6 & 3.5 & 4.0 & {\bf 4.9}\\
        \hline
        \hline
        \hline
        Overall ($\times10^{-7}$) & 6.7  & 11.97 & 13.96 & 19.39  & {\bf 19.89} \\
        \hline
        Space-Only ($\times10^{-3}$) & 3.5 & 6.8 & 6.7 & 6.6 & {\bf 6.9} \\
        \hline
        Space-Time ($\times10^{-7}$) & 8.02 & 15.87 & 19.00 & 18.84 & {\bf 20.44} \\
        \hline
        Space-Speed ($\times10^{-3}$) & 2.9 & 5.0 & 4.9 & 6.9 & 6.7\\
        \hline
    \end{tabular}
    \caption{Comparison on Forum (Top) and TrainStation (Bottom) based on AL metrics. {\bf Higher} is better. Numbers should only compared within the same row.)}
    \label{tab:AL_comparison}
\end{table}

\begin{figure*}[tb]
\centering
\includegraphics[width=\linewidth]{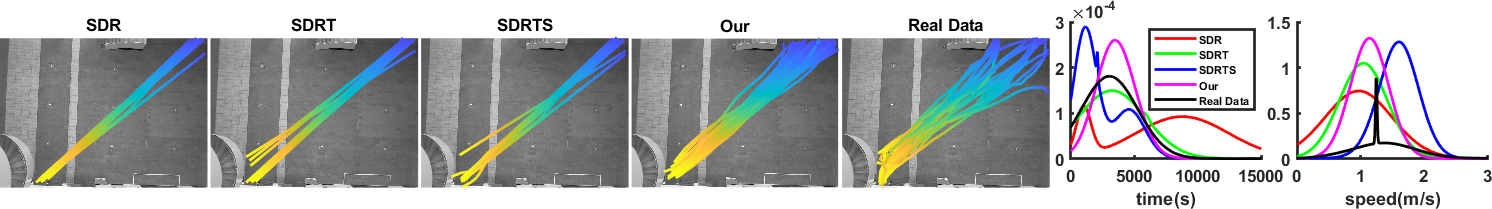}
\includegraphics[width=\linewidth]{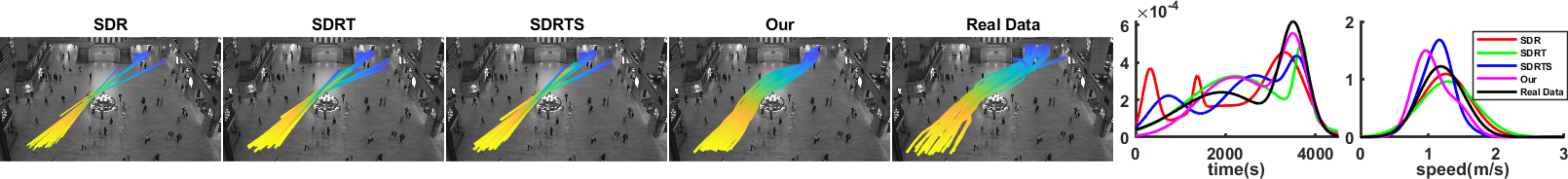}
\caption{Space flow P2 in Forum (Top) and P1 in TrainStation (Bottom) in different simulations. The y axes of the time and speed profiles indicate likelihood.}
\label{fig:spaceFlowSim}
\vspace{-1em}
\end{figure*}

Next, we show that it is possible to do more detailed comparisons using DPD metrics. Due to the space limit, we show one space flow from all simulation settings (\figref{spaceFlowSim}), and compare them in space only (DPD-Space), time only (DPD-Time) and time-speed (DPD-TS) in \tabref{DPD_trainStation}. In DPD-Space, all settings perform similarly because the space information is provided in all of them. In DPD-Time, SDRT \& SDRTS are better because they are both provided with the timing information. What is interesting is that SDRTS is worse than SDRT on the two flows in DPD-TS. Their main difference is that the desired speed in SDRTS is set to be the average speed of that trajectory, while the desired speed in SDRT is randomly drawn from a Gaussian estimated from real data. The latter achieves a slightly better performance on both flows in DPD-TS.

\begin{table}[tb]
    \centering
    \begin{tabular}{c|c|c|c|c}
    \hline
        Metric/Simulations &  SDR & SDRT & SDRTS & Ours\\
        \hline
        DPD-Space & 0.4751 & 0.3813 & 0.4374  & {\bf0.2988} \\
        \hline
        DPD-Time & 0.3545 & 0.0795 & 0.064 & {\bf0.0419}  \\
        \hline
        DPD-TS & 1.0 & 0.8879 & 1.0  & {\bf 0.4443} \\
        \hline
        \hline
        \hline
        DPD-Space &0.2753 &0.2461 &0.2423 & {\bf 0.1173} \\
        \hline
        DPD-Time & 0.0428 & 0.0319 & 0.0295 & {\bf 0.0213}  \\
        \hline
        DPD-TS &0.9970 &0.8157 &0.9724 & {\bf 0.5091} \\
        \hline
    \end{tabular}
    \caption{Comparison on space flow P2 in Forum (Top) and space flow P1 in TrainStation (Bottom) based on DPD metrics, both shown in \figref{std_vis}. {\bf Lower} is better.}
    \label{tab:DPD_trainStation}

\end{table}

Quantitative metrics for comparing simulated and real crowds have been proposed before. However, they either only compare individual motions \cite{guy_statistical_2012} or only space patterns \cite{wang_path_2016,wang_trending_2017}.  Holistically considering space, time \& speed has a combinatorial effect, leading to many explicable metrics evaluating different aspects of crowds (AL \& DPD metrics). This makes multi-faceted comparisons possible, which is unachievable in existing methods. Technically, the flexible design of THDP allows for different choices of marginalization, which greatly increases the evaluation versatility. This shows the theoretical superiority of THDP over existing methods.

\subsection{Guided Simulations}

Our automated simulation guidance proves to be superior to careful manual settings. We first show the AL results in \tabref{AL_comparison}. Our guided simulation outperforms all other settings that were carefully and manually set up. The superior performance is achieved in the Overall comparisons as well as most dimension-specific comparisons. Next, we show the same space flow of our guided simulation in \figref{spaceFlowSim}, in comparison with other settings. Qualitatively, SDR, SDRT and SDRTS generate narrower flows due to straight lines are simulated. In contrast, our simulation shows more realistic intra-flow randomness which led to a wider flow. It is much more similar to the real data. Quantitatively, we show the DPD results in \tabref{DPD_trainStation}. Again, our automated guidance outperforms all other settings.

Automated simulation guidance has only been attempted by a few researchers before \cite{wolinski_parameter_2014,karamouzas_crowd_2019}. However, their methods aim to guide simulators to reproduce low-level motions for the overall similarity with the data. Our approach aims to inform simulators with structured scene semantics. Moreover, it gives the freedom to the users so that the full semantics or partial semantics (e.g. the top n flows) can be used to simulate crowds, which no previous method can provide.

\subsection{Implementation Details}
For space discretization, we divide the image space of Forum, CarPark and TrainStation uniformly into $40 \times 40$, $40 \times 40$ and $120 \times 120$ pixel grids respectively. Since Forum is recorded by a top-down camera, we directly estimate the velocity from two consecutive observations in time. For CarPark and TrainStation, we estimate the velocity by reconstructing a top-down view via perspective projection. THDP also has hyper-parameters such as the scaling factors of every DP (totally 6 of them). Our inference method is not very sensitive to them because they are also sampled, as part of the CRFL sampling. Please refer to Appx. \ref{sec:APPHyperParameters} for details. In inference, we have a burn-in phase, during which we only use CRF on the Space-HDP and ignore the rest two HDPs. After the burn-in phase, we use CRFL on the full THDP. We found that it can greatly help the convergence of the inference. For crowd simulation, we use ORCA in Menge \cite{Curtis_menge_2015}.

We randomly select 664 trajectories in Forum, 1000 trajectories in CarPark and 1000 trajectories in Trainstation for performance tests. In each experiment, we split the data into segments in time domain to mimic fragmented video observations. The number of segments is a user-defined hyper-parameter and depends on the nature of the dataset. We chose the segment number to be 384, 87 and 28, for Forum, CarPark and TrainStation respectively to cover situations where the video is finely or roughly segmented. During training, we first run 5k CRF iterations on the Space-HDP only in the burn-in phase, then do the full CRFL on the whole THDP to speed up the mixing. After training, the numbers of space, time and speed modes are 25, 5 and 7 in Forum; 13, 6 and 6 in CarPark; 16, 3 and 4 in TrainStation. The training took 85.1, 11.5 and 7.8 minutes on Forum, Carpark and TrainStation, on a PC with an Intel i7-6700 3.4GHz CPU and 16GB memory.

\section{Discussion}
We chose MCMC to avoid the local minimum issue. (Stochastic) Variational Inference (VI) \cite{hoffman_stochastic_2013} and Geometric Optimization \cite{Yurochkin_geometric_2016} are theoretically faster. However, VI for a single HDP is already prone to local minimum \cite{wang_path_2016}. We also found the same issue with geometric optimization. Also, can we use three independent HDPs? Using independent HDPs essentially breaks the many-to-many associations between space, time and speed modes. It can cause mis-clustering due to that the clustering is done on different dimensions separately \cite{Wang_Globally_2016}.

The biggest limitation of our method does not consider the cross-scene transferability. Since the analysis focuses on the semantics in a given scene, it is unclear how the results can inspire simulation settings in unseen environments. In addition, our metrics do not directly reflect visual similarities on the individual level. We deliberately avoid the agent-level one-to-one comparison, to allow greater flexibility in simulation setting while maintaining statistical similarities. Also, we currently do not model high-level behaviors such as grouping, queuing, etc. This is due to that such information can only be obtained through human labelling which would incur massive workload and be therefore impractical on the chosen datasets. We intentionally chose unsupervised learning to deal with large datasets.

\section{Conclusions and Future Work}

In this paper, we present the first, to our best knowledge, multi-purpose framework for comprehensive crowd analysis, visualization, comparison (between real and simulated crowds) and simulation guidance. To this end, we proposed a new non-parametric Bayesian model called Triplet-HDP and a new inference method called Chinese Restaurant Franchise League. We have shown the effectiveness of our method on datasets varying in volume, duration, environment and crowd dynamics. 

In the future, we would like to extend the work to cross-environment prediction. It would be ideal if the modes learnt from given environments can be used to predict crowd behaviors in unseen environments. Preliminary results show that the semantics are tightly coupled with the layout of sub-spaces with designated functionalities. This means a subspace-functionality based semantic transfer is possible. Besides, we will look into using semi-supervised learning to identify and learn high level social behaviors, such as grouping and queuing.

\section*{Acknowledgement}
The project is partially supported by EPSRC (Ref:EP/R031193/1), the Fundamental Research Funds for the Central Universities (xzy012019048) and the National Natural Science Foundation of China (61602366).

\bibliographystyle{ACM-Reference-Format}
\bibliography{ref}


\begin{thebibliography}{42}


\ifx \showCODEN    \undefined \def \showCODEN     #1{\unskip}     \fi
\ifx \showDOI      \undefined \def \showDOI       #1{#1}\fi
\ifx \showISBNx    \undefined \def \showISBNx     #1{\unskip}     \fi
\ifx \showISBNxiii \undefined \def \showISBNxiii  #1{\unskip}     \fi
\ifx \showISSN     \undefined \def \showISSN      #1{\unskip}     \fi
\ifx \showLCCN     \undefined \def \showLCCN      #1{\unskip}     \fi
\ifx \shownote     \undefined \def \shownote      #1{#1}          \fi
\ifx \showarticletitle \undefined \def \showarticletitle #1{#1}   \fi
\ifx \showURL      \undefined \def \showURL       {\relax}        \fi
\providecommand\bibfield[2]{#2}
\providecommand\bibinfo[2]{#2}
\providecommand\natexlab[1]{#1}
\providecommand\showeprint[2][]{arXiv:#2}

\bibitem[\protect\citeauthoryear{Ali and Shah}{Ali and Shah}{2007}]%
        {ali2007lagrangian}
\bibfield{author}{\bibinfo{person}{Saad Ali} {and} \bibinfo{person}{Mubarak
  Shah}.} \bibinfo{year}{2007}\natexlab{}.
\newblock \showarticletitle{A lagrangian particle dynamics approach for crowd
  flow segmentation and stability analysis}. In \bibinfo{booktitle}{\emph{2007
  IEEE Conference on Computer Vision and Pattern Recognition}}. IEEE,
  \bibinfo{pages}{1--6}.
\newblock


\bibitem[\protect\citeauthoryear{Bian, Tian, Tang, and Tao}{Bian
  et~al\mbox{.}}{2018}]%
        {Bian_2018_survey}
\bibfield{author}{\bibinfo{person}{Jiang Bian}, \bibinfo{person}{Dayong Tian},
  \bibinfo{person}{Yuanyan Tang}, {and} \bibinfo{person}{Dacheng Tao}.}
  \bibinfo{year}{2018}\natexlab{}.
\newblock \showarticletitle{A survey on trajectory clustering analysis}.
\newblock \bibinfo{journal}{\emph{CoRR}}  \bibinfo{volume}{abs/1802.06971}
  (\bibinfo{year}{2018}).
\newblock
\showeprint[arxiv]{1802.06971}


\bibitem[\protect\citeauthoryear{Bishop}{Bishop}{2007}]%
        {bishop_pattern_2007}
\bibfield{author}{\bibinfo{person}{Christopher Bishop}.}
  \bibinfo{year}{2007}\natexlab{}.
\newblock \bibinfo{booktitle}{\emph{Pattern {Recognition} and {Machine}
  {Learning}}}.
\newblock \bibinfo{publisher}{Springer}, \bibinfo{address}{New York}.
\newblock


\bibitem[\protect\citeauthoryear{Chaker, Al~Aghbari, and Junejo}{Chaker
  et~al\mbox{.}}{2017}]%
        {chaker2017social}
\bibfield{author}{\bibinfo{person}{Rima Chaker}, \bibinfo{person}{Zaher
  Al~Aghbari}, {and} \bibinfo{person}{Imran~N Junejo}.}
  \bibinfo{year}{2017}\natexlab{}.
\newblock \showarticletitle{Social network model for crowd anomaly detection
  and localization}.
\newblock \bibinfo{journal}{\emph{Pattern Recognition}}  \bibinfo{volume}{61}
  (\bibinfo{year}{2017}), \bibinfo{pages}{266--281}.
\newblock


\bibitem[\protect\citeauthoryear{Charalambous, Karamouzas, Guy, and
  Chrysanthou}{Charalambous et~al\mbox{.}}{2014}]%
        {charalambous2014data}
\bibfield{author}{\bibinfo{person}{Panayiotis Charalambous},
  \bibinfo{person}{Ioannis Karamouzas}, \bibinfo{person}{Stephen~J Guy}, {and}
  \bibinfo{person}{Yiorgos Chrysanthou}.} \bibinfo{year}{2014}\natexlab{}.
\newblock \showarticletitle{A data-driven framework for visual crowd analysis}.
  In \bibinfo{booktitle}{\emph{Computer Graphics Forum}},
  Vol.~\bibinfo{volume}{33}. Wiley Online Library, \bibinfo{pages}{41--50}.
\newblock


\bibitem[\protect\citeauthoryear{Curtis, Best, and Manocha}{Curtis
  et~al\mbox{.}}{2016}]%
        {Curtis_menge_2015}
\bibfield{author}{\bibinfo{person}{Sean Curtis}, \bibinfo{person}{Andrew Best},
  {and} \bibinfo{person}{Dinesh Manocha}.} \bibinfo{year}{2016}\natexlab{}.
\newblock \showarticletitle{Menge: A Modular Framework for Simulating Crowd
  Movement}.
\newblock \bibinfo{journal}{\emph{Collective Dynamics}} \bibinfo{volume}{1},
  \bibinfo{number}{0} (\bibinfo{year}{2016}).
\newblock


\bibitem[\protect\citeauthoryear{Ennis, Peters, and O’Sullivan}{Ennis
  et~al\mbox{.}}{2011}]%
        {Ennis_perceptual_2011}
\bibfield{author}{\bibinfo{person}{Cathy Ennis}, \bibinfo{person}{Christopher
  Peters}, {and} \bibinfo{person}{Carol O’Sullivan}.}
  \bibinfo{year}{2011}\natexlab{}.
\newblock \showarticletitle{Perceptual Effects of Scene Context and Viewpoint
  for Virtual Pedestrian Crowds}.
\newblock \bibinfo{journal}{\emph{ACM Transaction on Applied Perception}}
  \bibinfo{volume}{8}, \bibinfo{number}{2}, Article \bibinfo{articleno}{10}
  (\bibinfo{date}{Feb.} \bibinfo{year}{2011}), \bibinfo{numpages}{22}~pages.
\newblock
\showISSN{1544-3558}


\bibitem[\protect\citeauthoryear{Ferguson}{Ferguson}{1973}]%
        {Ferguson_bayyesian_1973}
\bibfield{author}{\bibinfo{person}{Thomas~S. Ferguson}.}
  \bibinfo{year}{1973}\natexlab{}.
\newblock \showarticletitle{A Bayesian Analysis of Some Nonparametric
  Problems}.
\newblock \bibinfo{journal}{\emph{The Annals of Statistics}}
  \bibinfo{volume}{1}, \bibinfo{number}{2} (\bibinfo{year}{1973}),
  \bibinfo{pages}{209--230}.
\newblock
\showISSN{00905364}


\bibitem[\protect\citeauthoryear{Golas, Narain, and Lin}{Golas
  et~al\mbox{.}}{2013}]%
        {golas_hybrid_2013}
\bibfield{author}{\bibinfo{person}{Abhinav Golas}, \bibinfo{person}{Rahul
  Narain}, {and} \bibinfo{person}{Ming Lin}.} \bibinfo{year}{2013}\natexlab{}.
\newblock \showarticletitle{Hybrid {Long}-range {Collision} {Avoidance} for
  {Crowd} {Simulation}}. In \bibinfo{booktitle}{\emph{ACM SIGGRAPH Symposium on
  Interactive 3D Graphics and Games}}. \bibinfo{pages}{29--36}.
\newblock


\bibitem[\protect\citeauthoryear{Guy, van~den Berg, Liu, Lau, Lin, and
  Manocha}{Guy et~al\mbox{.}}{2012}]%
        {guy_statistical_2012}
\bibfield{author}{\bibinfo{person}{Stephen~J. Guy}, \bibinfo{person}{Jur
  van~den Berg}, \bibinfo{person}{Wenxi Liu}, \bibinfo{person}{Rynson Lau},
  \bibinfo{person}{Ming~C. Lin}, {and} \bibinfo{person}{Dinesh Manocha}.}
  \bibinfo{year}{2012}\natexlab{}.
\newblock \showarticletitle{A {Statistical} {Similarity} {Measure} for
  {Aggregate} {Crowd} {Dynamics}}.
\newblock \bibinfo{journal}{\emph{ACM Transaction on Graphics}}
  \bibinfo{volume}{31}, \bibinfo{number}{6} (\bibinfo{year}{2012}),
  \bibinfo{pages}{190:1--190:11}.
\newblock


\bibitem[\protect\citeauthoryear{Helbing et~al\mbox{.}}{Helbing
  et~al\mbox{.}}{1995}]%
        {Helbing_social_1995}
\bibfield{author}{\bibinfo{person}{Dirk Helbing} {et~al\mbox{.}}}
  \bibinfo{year}{1995}\natexlab{}.
\newblock \showarticletitle{Social Force Model for Pedestrian Dynamics}.
\newblock \bibinfo{journal}{\emph{Physical Review E}} (\bibinfo{year}{1995}).
\newblock


\bibitem[\protect\citeauthoryear{Hoffman, Blei, Wang, and Paisley}{Hoffman
  et~al\mbox{.}}{2013}]%
        {hoffman_stochastic_2013}
\bibfield{author}{\bibinfo{person}{Matthew~D. Hoffman},
  \bibinfo{person}{David~M. Blei}, \bibinfo{person}{Chong Wang}, {and}
  \bibinfo{person}{John Paisley}.} \bibinfo{year}{2013}\natexlab{}.
\newblock \showarticletitle{Stochastic {Variational} {Inference}}.
\newblock \bibinfo{journal}{\emph{Journal of Machine Learning Research}}
  \bibinfo{volume}{14}, \bibinfo{number}{1} (\bibinfo{year}{2013}),
  \bibinfo{pages}{1303--1347}.
\newblock


\bibitem[\protect\citeauthoryear{Jordao, Pettré, Christie, and Cani}{Jordao
  et~al\mbox{.}}{2014}]%
        {Jordao_crowd_2014}
\bibfield{author}{\bibinfo{person}{Kevin Jordao}, \bibinfo{person}{Julien
  Pettré}, \bibinfo{person}{Marc Christie}, {and} \bibinfo{person}{Marie-Paule
  Cani}.} \bibinfo{year}{2014}\natexlab{}.
\newblock \showarticletitle{{Crowd Sculpting: A Space-time Sculpting Method for
  Populating Virtual Environments}}.
\newblock \bibinfo{journal}{\emph{Computer Graphics Forum}}
  (\bibinfo{year}{2014}).
\newblock
\showISSN{1467-8659}


\bibitem[\protect\citeauthoryear{Karamouzas, Sohre, Hu, and Guy}{Karamouzas
  et~al\mbox{.}}{2018}]%
        {karamouzas_crowd_2019}
\bibfield{author}{\bibinfo{person}{Ioannis Karamouzas}, \bibinfo{person}{Nick
  Sohre}, \bibinfo{person}{Ran Hu}, {and} \bibinfo{person}{Stephen~J. Guy}.}
  \bibinfo{year}{2018}\natexlab{}.
\newblock \showarticletitle{Crowd Space: A Predictive Crowd Analysis
  Technique}.
\newblock \bibinfo{journal}{\emph{ACM Transaction on Graphics}}
  \bibinfo{volume}{37}, \bibinfo{number}{6}, Article \bibinfo{articleno}{186}
  (\bibinfo{date}{Dec.} \bibinfo{year}{2018}), \bibinfo{numpages}{14}~pages.
\newblock
\showISSN{0730-0301}


\bibitem[\protect\citeauthoryear{Kauffman and Rousseeuw}{Kauffman and
  Rousseeuw}{2005}]%
        {Kaufman_2005_introduction}
\bibfield{author}{\bibinfo{person}{Leonard Kauffman} {and}
  \bibinfo{person}{Peter~J. Rousseeuw}.} \bibinfo{year}{2005}\natexlab{}.
\newblock \bibinfo{booktitle}{\emph{Finding Groups in Data: An Introduction to
  Cluster Analysis}}.
\newblock \bibinfo{publisher}{John Wiley \& Sons}.
\newblock


\bibitem[\protect\citeauthoryear{Lee, Choi, Hong, and Lee}{Lee
  et~al\mbox{.}}{2007}]%
        {lee2007group}
\bibfield{author}{\bibinfo{person}{Kang~Hoon Lee}, \bibinfo{person}{Myung~Geol
  Choi}, \bibinfo{person}{Qyoun Hong}, {and} \bibinfo{person}{Jehee Lee}.}
  \bibinfo{year}{2007}\natexlab{}.
\newblock \showarticletitle{Group behavior from video: a data-driven approach
  to crowd simulation}. In \bibinfo{booktitle}{\emph{Proceedings of the 2007
  ACM SIGGRAPH/Eurographics symposium on Computer animation}}.
  \bibinfo{pages}{109--118}.
\newblock


\bibitem[\protect\citeauthoryear{Lemercier, Jelic, Kulpa, Hua, Fehrenbach,
  Degond, Appert-Rolland, Donikian, and Pettré}{Lemercier
  et~al\mbox{.}}{2012}]%
        {lemercier_realistic_2012}
\bibfield{author}{\bibinfo{person}{S. Lemercier}, \bibinfo{person}{A. Jelic},
  \bibinfo{person}{R. Kulpa}, \bibinfo{person}{J. Hua}, \bibinfo{person}{J.
  Fehrenbach}, \bibinfo{person}{P. Degond}, \bibinfo{person}{C.
  Appert-Rolland}, \bibinfo{person}{S. Donikian}, {and} \bibinfo{person}{J.
  Pettré}.} \bibinfo{year}{2012}\natexlab{}.
\newblock \showarticletitle{Realistic {Following} {Behaviors} for {Crowd}
  {Simulation}}.
\newblock \bibinfo{journal}{\emph{Computer Graphics Forum}}
  \bibinfo{volume}{31}, \bibinfo{number}{2} (\bibinfo{year}{2012}),
  \bibinfo{pages}{489--498}.
\newblock


\bibitem[\protect\citeauthoryear{Lerner, Chrysanthou, Shamir, and
  Cohen-Or}{Lerner et~al\mbox{.}}{2009}]%
        {lerner2009data}
\bibfield{author}{\bibinfo{person}{Alon Lerner}, \bibinfo{person}{Yiorgos
  Chrysanthou}, \bibinfo{person}{Ariel Shamir}, {and} \bibinfo{person}{Daniel
  Cohen-Or}.} \bibinfo{year}{2009}\natexlab{}.
\newblock \showarticletitle{Data driven evaluation of crowds}. In
  \bibinfo{booktitle}{\emph{International Workshop on Motion in Games}}.
  Springer, \bibinfo{pages}{75--83}.
\newblock


\bibitem[\protect\citeauthoryear{López, Chaumette, Marchand, and
  Pettré}{López et~al\mbox{.}}{2019}]%
        {Lopez_character_2019}
\bibfield{author}{\bibinfo{person}{A López}, \bibinfo{person}{F Chaumette},
  \bibinfo{person}{E Marchand}, {and} \bibinfo{person}{J Pettré}.}
  \bibinfo{year}{2019}\natexlab{}.
\newblock \showarticletitle{Character navigation in dynamic environments based
  on optical flow}. In \bibinfo{booktitle}{\emph{Proceedings of Eurographics
  2019}} \emph{(\bibinfo{series}{Eurographics 2019})}.
  \bibinfo{publisher}{Eurographics}.
\newblock


\bibitem[\protect\citeauthoryear{Lu et~al\mbox{.}}{Lu et~al\mbox{.}}{2019}]%
        {Liu_ADCrowdNet_2019}
\bibfield{author}{\bibinfo{person}{Ning Lu} {et~al\mbox{.}}}
  \bibinfo{year}{2019}\natexlab{}.
\newblock \showarticletitle{ADCrowdNet: An Attention-injective Deformable
  Convolutional Networkfor Crowd Understanding}.
\newblock \bibinfo{journal}{\emph{IEEE Conference on Computer Vision and
  Pattern Recognition}} (\bibinfo{year}{2019}).
\newblock


\bibitem[\protect\citeauthoryear{Majecka}{Majecka}{2009}]%
        {majecka_statistical_2009}
\bibfield{author}{\bibinfo{person}{B. Majecka}.}
  \bibinfo{year}{2009}\natexlab{}.
\newblock \emph{\bibinfo{title}{Statistical models of pedestrian behaviour in
  the {Forum}}}.
\newblock {MSc} {Dissertation}. \bibinfo{school}{School of Informatics,
  University of Edinburgh}, \bibinfo{address}{Edinburgh}.
\newblock


\bibitem[\protect\citeauthoryear{Mehran, Oyama, and Shah}{Mehran
  et~al\mbox{.}}{2009}]%
        {mehran2009abnormal}
\bibfield{author}{\bibinfo{person}{Ramin Mehran}, \bibinfo{person}{Alexis
  Oyama}, {and} \bibinfo{person}{Mubarak Shah}.}
  \bibinfo{year}{2009}\natexlab{}.
\newblock \showarticletitle{Abnormal crowd behavior detection using social
  force model}. In \bibinfo{booktitle}{\emph{2009 IEEE Conference on Computer
  Vision and Pattern Recognition}}. IEEE, \bibinfo{pages}{935--942}.
\newblock


\bibitem[\protect\citeauthoryear{Narain, Golas, Curtis, and Lin}{Narain
  et~al\mbox{.}}{2009}]%
        {narain_aggregate_2009}
\bibfield{author}{\bibinfo{person}{Rahul Narain}, \bibinfo{person}{Abhinav
  Golas}, \bibinfo{person}{Sean Curtis}, {and} \bibinfo{person}{Ming~C. Lin}.}
  \bibinfo{year}{2009}\natexlab{}.
\newblock \showarticletitle{Aggregate {Dynamics} for {Dense} {Crowd}
  {Simulation}}.
\newblock \bibinfo{journal}{\emph{ACM Transaction on Graphics}}
  \bibinfo{volume}{28}, \bibinfo{number}{5} (\bibinfo{year}{2009}),
  \bibinfo{pages}{122:1--122:8}.
\newblock


\bibitem[\protect\citeauthoryear{Rasmussen}{Rasmussen}{1999}]%
        {Rasmussen_infinite_1999}
\bibfield{author}{\bibinfo{person}{Carl~Edward Rasmussen}.}
  \bibinfo{year}{1999}\natexlab{}.
\newblock \showarticletitle{The Infinite Gaussian Mixture Model}. In
  \bibinfo{booktitle}{\emph{International Conference on Neural Information
  Processing Systems}} (Denver, CO) \emph{(\bibinfo{series}{NIPS’99})}.
  \bibinfo{publisher}{MIT Press}, \bibinfo{address}{Cambridge, MA, USA},
  \bibinfo{pages}{554–560}.
\newblock


\bibitem[\protect\citeauthoryear{Ren, Xiang, Xiao, Yang, Manocha, and Jin}{Ren
  et~al\mbox{.}}{2018}]%
        {Ren_heter_2018}
\bibfield{author}{\bibinfo{person}{Jiaping Ren}, \bibinfo{person}{Wei Xiang},
  \bibinfo{person}{Yangxi Xiao}, \bibinfo{person}{Ruigang Yang},
  \bibinfo{person}{Dinesh Manocha}, {and} \bibinfo{person}{Xiaogang Jin}.}
  \bibinfo{year}{2018}\natexlab{}.
\newblock \showarticletitle{Heter-Sim: Heterogeneous multi-agent systems
  simulation by interactive data-driven optimization}.
\newblock \bibinfo{journal}{\emph{CoRR}}  \bibinfo{volume}{abs/1812.00307}
  (\bibinfo{year}{2018}).
\newblock
\showeprint[arxiv]{1812.00307}


\bibitem[\protect\citeauthoryear{Ren, Charalambous, Bruneau, Peng, and
  Pettré}{Ren et~al\mbox{.}}{2016}]%
        {Ren_group_2016}
\bibfield{author}{\bibinfo{person}{Zeng Ren}, \bibinfo{person}{P.
  Charalambous}, \bibinfo{person}{J. Bruneau}, \bibinfo{person}{Q. Peng}, {and}
  \bibinfo{person}{J. Pettré}.} \bibinfo{year}{2016}\natexlab{}.
\newblock \showarticletitle{Group modelling: A unified velocity-based
  approach}.
\newblock \bibinfo{journal}{\emph{Computer Graphics Forum}}
  (\bibinfo{year}{2016}).
\newblock


\bibitem[\protect\citeauthoryear{Sabokrou et~al\mbox{.}}{Sabokrou
  et~al\mbox{.}}{2017}]%
        {Sabokrou_deep_2019}
\bibfield{author}{\bibinfo{person}{Mohammad Sabokrou} {et~al\mbox{.}}}
  \bibinfo{year}{2017}\natexlab{}.
\newblock \showarticletitle{Deep-cascade:cascading 3D deep neural networks for
  fast anomaly detection and localization in crowded scenes}.
\newblock \bibinfo{journal}{\emph{IEEE Transaction on Image Processing}}
  (\bibinfo{year}{2017}).
\newblock


\bibitem[\protect\citeauthoryear{Sha, Lucey, Yue, Wei, Hobbs, Rohlf, and
  Sridharan}{Sha et~al\mbox{.}}{2018}]%
        {sha2018interactive}
\bibfield{author}{\bibinfo{person}{Long Sha}, \bibinfo{person}{Patrick Lucey},
  \bibinfo{person}{Yisong Yue}, \bibinfo{person}{Xinyu Wei},
  \bibinfo{person}{Jennifer Hobbs}, \bibinfo{person}{Charlie Rohlf}, {and}
  \bibinfo{person}{Sridha Sridharan}.} \bibinfo{year}{2018}\natexlab{}.
\newblock \showarticletitle{Interactive sports analytics: An intelligent
  interface for utilizing trajectories for interactive sports play retrieval
  and analytics}.
\newblock \bibinfo{journal}{\emph{ACM Transactions on Computer-Human
  Interaction (TOCHI)}} \bibinfo{volume}{25}, \bibinfo{number}{2}
  (\bibinfo{year}{2018}), \bibinfo{pages}{1--32}.
\newblock


\bibitem[\protect\citeauthoryear{Sha, Lucey, Zheng, Kim, Yue, and
  Sridharan}{Sha et~al\mbox{.}}{2017}]%
        {sha2017fine}
\bibfield{author}{\bibinfo{person}{Long Sha}, \bibinfo{person}{Patrick Lucey},
  \bibinfo{person}{Stephan Zheng}, \bibinfo{person}{Taehwan Kim},
  \bibinfo{person}{Yisong Yue}, {and} \bibinfo{person}{Sridha Sridharan}.}
  \bibinfo{year}{2017}\natexlab{}.
\newblock \showarticletitle{Fine-grained retrieval of sports plays using
  tree-based alignment of trajectories}.
\newblock  (\bibinfo{year}{2017}).
\newblock
\showeprint[arxiv]{1710.02255}


\bibitem[\protect\citeauthoryear{Shen, Henry, Wang, Ho, Komura, and Shum}{Shen
  et~al\mbox{.}}{2018}]%
        {Shen_data_2018}
\bibfield{author}{\bibinfo{person}{Yijun Shen}, \bibinfo{person}{Joseph Henry},
  \bibinfo{person}{He Wang}, \bibinfo{person}{Edmond S.~L. Ho},
  \bibinfo{person}{Taku Komura}, {and} \bibinfo{person}{Hubert P.~H. Shum}.}
  \bibinfo{year}{2018}\natexlab{}.
\newblock \showarticletitle{Data-Driven Crowd Motion Control With Multi-Touch
  Gestures}.
\newblock \bibinfo{journal}{\emph{Computer Graphics Forum}}
  \bibinfo{volume}{37}, \bibinfo{number}{6} (\bibinfo{year}{2018}),
  \bibinfo{pages}{382--394}.
\newblock
\showeprint{https://onlinelibrary.wiley.com/doi/pdf/10.1111/cgf.13333}


\bibitem[\protect\citeauthoryear{Shi and {Malik}}{Shi and {Malik}}{2000}]%
        {Shi_2000_normalized}
\bibfield{author}{\bibinfo{person}{Jianbo Shi} {and} \bibinfo{person}{J.
  {Malik}}.} \bibinfo{year}{2000}\natexlab{}.
\newblock \showarticletitle{Normalized cuts and image segmentation}.
\newblock \bibinfo{journal}{\emph{IEEE Transactions on Pattern Analysis and
  Machine Intelligence}} \bibinfo{volume}{22}, \bibinfo{number}{8}
  (\bibinfo{year}{2000}), \bibinfo{pages}{888--905}.
\newblock


\bibitem[\protect\citeauthoryear{Teh, Jordan, Beal, and Blei}{Teh
  et~al\mbox{.}}{2006}]%
        {teh_hierarchical_2006}
\bibfield{author}{\bibinfo{person}{Yee~Whye Teh}, \bibinfo{person}{Michael~I.
  Jordan}, \bibinfo{person}{Matthew~J. Beal}, {and} \bibinfo{person}{David~M.
  Blei}.} \bibinfo{year}{2006}\natexlab{}.
\newblock \showarticletitle{Hierarchical {Dirichlet} {Processes}}.
\newblock \bibinfo{journal}{\emph{Journal of American Statistical Association}}
  \bibinfo{volume}{101}, \bibinfo{number}{476} (\bibinfo{year}{2006}),
  \bibinfo{pages}{1566--1581}.
\newblock


\bibitem[\protect\citeauthoryear{van~den Berg, Lin, and Manocha}{van~den Berg
  et~al\mbox{.}}{2008}]%
        {Berg_Reciprocal_2008}
\bibfield{author}{\bibinfo{person}{J. van~den Berg}, \bibinfo{person}{Ming~C.
  Lin}, {and} \bibinfo{person}{Dinesh Manocha}.}
  \bibinfo{year}{2008}\natexlab{}.
\newblock \showarticletitle{Reciprocal Velocity Obstacles for real-time
  multi-agent navigation}.
\newblock \bibinfo{journal}{\emph{IEEE International Conference on Robotics and
  Automation}} (\bibinfo{year}{2008}).
\newblock


\bibitem[\protect\citeauthoryear{Wang, Ondřej, and O'Sullivan}{Wang
  et~al\mbox{.}}{2016}]%
        {wang_path_2016}
\bibfield{author}{\bibinfo{person}{He Wang}, \bibinfo{person}{Jan Ondřej},
  {and} \bibinfo{person}{Carol O'Sullivan}.} \bibinfo{year}{2016}\natexlab{}.
\newblock \showarticletitle{Path {Patterns}: {Analyzing} and {Comparing} {Real}
  and {Simulated} {Crowds}}. In \bibinfo{booktitle}{\emph{Proceedings of the
  20th {ACM} {SIGGRAPH} {Symposium} on {Interactive} 3D {Graphics} and
  {Games}}} \emph{(\bibinfo{series}{I3D '16})}. \bibinfo{publisher}{ACM},
  \bibinfo{address}{New York, NY, USA}, \bibinfo{pages}{49--57}.
\newblock
\showISBNx{978-1-4503-4043-4}
\urldef\tempurl%
\url{https://doi.org/10.1145/2856400.2856410}
\showDOI{\tempurl}


\bibitem[\protect\citeauthoryear{Wang, Ondřej, and O'Sullivan}{Wang
  et~al\mbox{.}}{2017}]%
        {wang_trending_2017}
\bibfield{author}{\bibinfo{person}{He Wang}, \bibinfo{person}{Jan Ondřej},
  {and} \bibinfo{person}{Carol O'Sullivan}.} \bibinfo{year}{2017}\natexlab{}.
\newblock \showarticletitle{Trending Paths: A New Semantic-level Metric for
  Comparing Simulated and Real Crowd Data}.
\newblock \bibinfo{journal}{\emph{IEEE Transactions on Visualization and
  Computer Graphics}} \bibinfo{volume}{23}, \bibinfo{number}{5}
  (\bibinfo{year}{2017}), \bibinfo{pages}{1454--1464}.
\newblock
\showISSN{1077-2626}


\bibitem[\protect\citeauthoryear{Wang and O'Sullivan}{Wang and
  O'Sullivan}{2016}]%
        {Wang_Globally_2016}
\bibfield{author}{\bibinfo{person}{He Wang} {and} \bibinfo{person}{Carol
  O'Sullivan}.} \bibinfo{year}{2016}\natexlab{}.
\newblock \bibinfo{booktitle}{\emph{Globally Continuous and Non-Markovian Crowd
  Activity Analysis from Videos}}.
\newblock \bibinfo{publisher}{Springer International Publishing},
  \bibinfo{address}{Cham}, \bibinfo{pages}{527--544}.
\newblock
\showISBNx{978-3-319-46454-1}


\bibitem[\protect\citeauthoryear{Wang et~al\mbox{.}}{Wang
  et~al\mbox{.}}{2019}]%
        {Wang_learning_2019}
\bibfield{author}{\bibinfo{person}{Qi Wang} {et~al\mbox{.}}}
  \bibinfo{year}{2019}\natexlab{}.
\newblock \showarticletitle{Learning from Synthetic Data for Crowd Counting in
  the Wild}.
\newblock \bibinfo{journal}{\emph{IEEE Conference on Computer Vision and
  Pattern Recognition}} (\bibinfo{year}{2019}).
\newblock


\bibitem[\protect\citeauthoryear{Wang, Ma, Ng, and {Grimson}}{Wang
  et~al\mbox{.}}{2008}]%
        {wang_trajectory_2008}
\bibfield{author}{\bibinfo{person}{Xiaogang Wang}, \bibinfo{person}{Keng~Teck
  Ma}, \bibinfo{person}{Gee-Wah Ng}, {and} \bibinfo{person}{W.~E.~L.
  {Grimson}}.} \bibinfo{year}{2008}\natexlab{}.
\newblock \showarticletitle{Trajectory analysis and semantic region modeling
  using a nonparametric Bayesian model}. In \bibinfo{booktitle}{\emph{IEEE
  Conference on Computer Vision and Pattern Recognition}}.
  \bibinfo{pages}{1--8}.
\newblock
\showISSN{1063-6919}


\bibitem[\protect\citeauthoryear{Wolinski, Guy, Olivier, Lin, Manocha, and
  Pettré}{Wolinski et~al\mbox{.}}{2014}]%
        {wolinski_parameter_2014}
\bibfield{author}{\bibinfo{person}{David Wolinski}, \bibinfo{person}{Stephen~J.
  Guy}, \bibinfo{person}{Anne-Hélène Olivier}, \bibinfo{person}{Ming~C. Lin},
  \bibinfo{person}{Dinesh Manocha}, {and} \bibinfo{person}{Julien Pettré}.}
  \bibinfo{year}{2014}\natexlab{}.
\newblock \showarticletitle{Parameter estimation and comparative evaluation of
  crowd simulations}.
\newblock \bibinfo{journal}{\emph{Computer Graphics Forum}}
  \bibinfo{volume}{33}, \bibinfo{number}{2} (\bibinfo{year}{2014}),
  \bibinfo{pages}{303--312}.
\newblock


\bibitem[\protect\citeauthoryear{Xu et~al\mbox{.}}{Xu et~al\mbox{.}}{2018}]%
        {Xu_encoding_2018}
\bibfield{author}{\bibinfo{person}{Yanyu Xu} {et~al\mbox{.}}}
  \bibinfo{year}{2018}\natexlab{}.
\newblock \showarticletitle{Encoding Crowd Interaction with Deep Neural Network
  for Pedestrian Trajectory Prediction}.
\newblock \bibinfo{journal}{\emph{IEEE Conference on Computer Vision and
  Pattern Recognition}} (\bibinfo{year}{2018}).
\newblock


\bibitem[\protect\citeauthoryear{{Yi}, {Li}, and {Wang}}{{Yi}
  et~al\mbox{.}}{2015}]%
        {Yi_understanding_2015}
\bibfield{author}{\bibinfo{person}{S. {Yi}}, \bibinfo{person}{H. {Li}}, {and}
  \bibinfo{person}{X. {Wang}}.} \bibinfo{year}{2015}\natexlab{}.
\newblock \showarticletitle{Understanding pedestrian behaviors from stationary
  crowd groups}. In \bibinfo{booktitle}{\emph{IEEE Conference on Computer
  Vision and Pattern Recognition}}. \bibinfo{pages}{3488--3496}.
\newblock
\showISSN{1063-6919}


\bibitem[\protect\citeauthoryear{Yurochkin and Nguyen}{Yurochkin and
  Nguyen}{2016}]%
        {Yurochkin_geometric_2016}
\bibfield{author}{\bibinfo{person}{Mikhail Yurochkin} {and}
  \bibinfo{person}{XuanLong Nguyen}.} \bibinfo{year}{2016}\natexlab{}.
\newblock \showarticletitle{Geometric Dirichlet Means Algorithm for topic
  inference}. In \bibinfo{booktitle}{\emph{International Conference on Neural
  Information Processing Systems}}.
\newblock


\end{thebibliography}

\appendix
\section{Chinese Restaurant Franchise}
\label{sec:AppCRF}
To give the mathematical derivation of the sampling process described in \secref{CRF}, we first give meanings to the variables in \figref{models} Left. $\theta_{ji}$ is the dish choice made by $x_{ji}$, the $i$th customer in the $j$th restaurant. $G_j$ is the tables with dishes and the dishes are from the global menu $G$. Since $\theta_{ji}$ indicates the choice of tables and therefore dishes, we use some auxiliary variables to represent the process. We introduce $t_{ji}$ and $k_{jt}$ as the indices of the table and the dish on the table chosen by $x_{ji}$. We also denote $m_{jk}$ as the number of tables serving the $k$th dish in restaurant $j$ and $n_{jtk}$ as the number of customers at table $t$ in restaurant $j$ having the $k$th dish. We also use them to represent accumulative indicators such as $m_{\cdot k}$ representing the total number of tables serving the $k$th dish. We also use superscript to indicate which customer or table is removed. If customer $x_{ji}$ is removed, then $n_{jtk}^{-ji}$ is the number of customers at the table $t$ in restaurant $j$ having the $k$th dish without the customer $x_{ji}$.

{\bf Customer-level sampling}. To choose a table for $x_{ji}$ (line 5 in \algref{CRF}), we sample a table index $t_{ji}$:
\begin{equation}
\label{eq:tableSampling}
p(t_{ji} = t | \mathbf{t^{-ji}}, \mathbf{k}) \propto 
\begin{cases}
& n_{jt\cdot}^{-ji} f_{k_{jt}}^{-x_{ji}}(x_{ji})\ \text{if $t$ already exists} \\
&\alpha_j p(x_{ji}|\mathbf{t^{-ji}}, t_{ji} = t^{new}, \mathbf{k})\ \text{if $t = t^{new}$}
\end{cases}
\end{equation}
where $n_{jt\cdot}^{-ji}$ is the number of customers at table $t$ (table popularity), and $f_{k_{jt}}^{-x_{ji}}(x_{ji})$ is how much $x_{ji}$ likes the $k_{jt}$th dish, $f_{k_{jt}}$, served on that table (dish preference). $f_{k_{jt}}$ is the dish and thus is a problem-specific probability distribution. $f_{k_{jt}}^{-x_{ji}}(x_{ji})$ is the likelihood of $x_{ji}$ on $f_{k_{jt}}$. In our problem, $f_{k_{jt}}$ is Multinomial if it is the Space-HDP or otherwise Normal. $\alpha_j$ is the parameter in \eqref{HDP}, so it controls how likely $x_{ji}$ will create a new table, after which she needs to choose a dish according to $p(x_{ji}|\mathbf{t^{-ji}}, t_{ji} = t^{new}, \mathbf{k})$. When a new table is created, $t_{ji} = t^{new}$, we need sampling a dish (line 7 in \algref{CRF}), indexed by $k_{jt^{new}}$, according to:

\begin{equation}
\label{eq:dishSampling}
p(k_{jt^{new}} = k | \mathbf{t}, \mathbf{k^{-jt^{new}}}) \propto 
\begin{cases}
& m_{\cdot k} f_{k}^{-x_{ji}}(x_{ji})\ \text{if $k$ already exists} \\
&\gamma f_{k^{new}}^{-x_{ji}}(x_{ji})\ \text{if $k = k^{new}$}
\end{cases}
\end{equation}
where $m_{\cdot k}$ is the total number of tables across all restaurants serving the $k$th dish (dish popularity). $f_{k}^{-x_{ji}}(x_{ji})$ is how much $x_{ji}$ like the $k$th dish, again the likelihood of $x_{ji}$ on $f_k$. $\gamma$ is the parameter in \eqref{HDP}, so it controls how likely a new dish will be created.

{\bf Table-level sampling}. Next we sample a dish for a table (line 11 in \algref{CRF}). We denote all customers at the $t$th table in the $j$th restaurant as $\mathbf{x_{jt}}$. Then we sample its dish $k_{jt}$ according to:

\begin{equation}
\label{eq:tableDishSampling}
p(k_{jt} = k | \mathbf{t}, \mathbf{k^{-jt}}) \propto 
\begin{cases}
& m_{\cdot k}^{-jt} f_{k}^{\mathbf{-x_{jt}}}(\mathbf{x_{jt}})\ \text{if $k$ already exists} \\
&\gamma f_{k^{new}}^{\mathbf{-x_{jt}}}(\mathbf{x_{jt}})\ \text{if $k = k^{new}$}
\end{cases}
\end{equation}
Similarly, $m_{\cdot k}^{-jt}$ is the total number of tables across all restaurants serving the $k$th dish, without $\mathbf{x_{jt}}$ (dish popularity). $f_{k}^{\mathbf{-x_{jt}}}(\mathbf{x_{jt}})$ is how much the group of customers $\mathbf{x_{jt}}$ likes the $k$th dish (dish preference). This time, $f_{k}^{\mathbf{-x_{jt}}}(\mathbf{x_{jt}})$ is a joint probability of all $x_{ji} \in \mathbf{x_{jt}}$.

Finally, in both \eqref{dishSampling} and \eqref{tableDishSampling}, we need to sample a new dish. This is done by sampling a new distribution from the base distribution $H$, $\phi_k \sim H$. After inference, the weights \bm{$\beta$} can be computed as $\bm{\beta} \sim Dirichlet(m_{\cdot 1}, m_{\cdot 2}, \cdots, m_{\cdot k}, \gamma)$. The choice of $H$ is related to the data. In our metaphor, the dishes of the Space-HDP are flows so we use Dirichlet. In the Time-HDP and Speed-HDP, the dishes are modes of time and speed which are Normals. So we use Normal-Inverse-Gamma for $H$. The choices are because Dirchlet and Norma-Inverse-Gamma are the \textit{conjugate priors} of Multinomial and Normal respectively. The whole CRF sampling is done by iteratively computing \eqref{tableSampling} to \eqref{tableDishSampling}. The dish number will dynamically increase/decrease until the sampling mixes. In this way, we do not need to know in advance how many space flows or time modes or speed modes there are because they will be automatically learnt.

\section{Chinese Restaurant Franchise League}
\label{sec:AppCRFL}
\subsection{Customer Level Sampling}
When we do customer-level sampling to sample a new table (line 8 in \algref{CRFL}), the left side of \eqref{tableSampling} becomes:
\begin{equation}
\label{eq:AppCRFLTable}
p(t_{ji} = t, x_{ji}, y_{kd}, z_{kc} | \mathbf{x^{-ji}}, \mathbf{t^{-ji}}, \mathbf{k}, \mathbf{y^{-kd}}, \mathbf{o^{-kd}}, \mathbf{l}, \mathbf{z^{-kc}}, \mathbf{p^{-kc}}, \mathbf{q}) 
\end{equation}
So whether $y_{kd}$ and $z_{kc}$ like the new restaurants should be taken into consideration. After applying Bayesian rules and factorization on \eqref{AppCRFLTable}, we have:

\begin{align}
\label{eq:AppCRFLTableDe}
p(t_{ji} = t, x_{ji}, &y_{kd}, z_{kc} | \bullet) = p(t_{ji} | \mathbf{t^{-ji}}, \mathbf{k}) \nonumber \\
&p(x_{ji} | y_{kd}, z_{kc}, t_{ji} = t, k_{jt} = k, \bullet) \nonumber \\
&p(y_{kd} | t_{ji} = t, k_{jt} = k, \mathbf{y^{-kd}}, \mathbf{o^{-kd}}, \mathbf{l}) \nonumber \\
&p(z_{kc} | t_{ji} = t, k_{jt} = k, \mathbf{z^{-kc}}, \mathbf{p^{-kc}}, \mathbf{q})
\end{align}
where $\bullet$ is \{$\mathbf{x^{-ji}}, \mathbf{t^{-ji}}, \mathbf{k}, \mathbf{y^{-kd}}, \mathbf{o^{-kd}}, \mathbf{l}, \mathbf{z^{-kc}}, \mathbf{p^{-kc}}, \mathbf{q}$\}. The four probabilities on the right-hand side of \eqref{AppCRFLTableDe} have intuitive meanings. $p(t_{ji} | \mathbf{t^{-ji}}, \mathbf{k})$ and $p(x_{ji} | y_{kd}, z_{kc}, t_{ji} = t, k_{jt} = k, \bullet)$ are the table popularity and dish preference of $x_{ji}$ in the space-HDP:

\begin{equation}
\label{eq:AppCRFLTable1}
p(t_{ji} | \mathbf{t^{-ji}}, \mathbf{k}) \propto
\begin{cases}
& n_{jt}^{-ji}\ \text{if $t$ already exists} \\
& \alpha_j\ \text{if $t = t^{new}$}
\end{cases}
\end{equation}

\begin{equation}
\label{eq:AppCRFLTable2}
p(x_{ji} | y_{kd}, z_{kc}, t_{ji} = t, k_{jt} = k, \bullet) \propto
\begin{cases}
&f_{k_{jt}}^{-x_{ji}}(x_{ji})\ \text{if $t$ exists} \\
&m_{\cdot k} f_{k}^{-x_{ji}}(x_{ji})\ \text{else if $k$ exists} \\
&\gamma f_{k^{new}}^{-x_{ji}}(x_{ji})\ \text{if $k = k^{new}$}
\end{cases}
\end{equation}
\eqref{AppCRFLTable1} and \eqref{AppCRFLTable2} are just re-organization of \eqref{tableSampling} and \eqref{dishSampling}. The remaining $p(y_{kd} | t_{ji} = t, k_{jt} = k, \mathbf{y^{-kd}}, \mathbf{o^{-kd}}, \mathbf{l})$ and $p(z_{kc} | t_{ji} = t, k_{jt} = k, \mathbf{z^{-kc}}, \mathbf{p^{-kc}}, \mathbf{q})$ can be seen as how much the time-customer $y_{kd}$ and speed-customer $z_{kc}$ like the $k$th time and speed restaurant respectively (restaurant preference). This restaurant preference does not appear in single HDPs and thus need special treatment. This is the first major difference between CRFL and CRF. Since we propose the same treatment for both, we only explain the time-restaurant preference treatment here.

If every time we sample a $t_{ji}$, we compute $p(y_{kd} | t_{ji} = t, k_{jt} = k, \mathbf{y^{-kd}}, \mathbf{o^{-kd}}, \mathbf{l})$ on every time table in every time-restaurant, it will be prohibitively slow. We therefore marginalize over all the time tables in a time-restaurant, to get a general restaurant preference of $y_{kd}$:

\begin{align}
\label{eq:AppCRFLTable3}
p(y_{kd} | t_{ji} = t, k_{jt} = k, \mathbf{y^{-kd}}, \mathbf{o^{-kd}}, \mathbf{l}) = \nonumber\\
\sum_{o_{kd} = 1}^{h_{k\cdot}}p(o_{kd} = o | t_{ji} = t, k_{jt} = k, \mathbf{y^{-kd}}, \mathbf{o^{-kd}}) \nonumber \\
p(y_{kd} | o_{kd} = o, l_{ko} = l, \mathbf{l})
\end{align}
where $o_{kd}$ is the table choice of $y_{kd}$ in the $kth$ time-restaurant. $l_{ko}$ is the time-dish served on the $o$th table in the $k$th time-restaurant.$h_{k\cdot}$ is the total number of tables in the $k$th time-restaurant. Similar to \eqref{AppCRFLTable1} and \eqref{AppCRFLTable2}:

\begin{equation}
\label{eq:AppCRFLTable4}
p(o_{kd} = o | t_{ji} = t, k_{jt} = k, \mathbf{y^{-kd}}, \mathbf{o^{-kd}}) \propto
\begin{cases}
& s_{ko}^{-kd}\ \text{if $o$ exists} \\
& \epsilon_k\ \text{if $o_{kd} = o^{new}$}
\end{cases}
\end{equation}
where $s_{ko}^{-kd}$ is the number of time-customers already at the $o$th table and $\epsilon_k$ is the scaling factor.

\begin{equation}
\label{eq:AppCRFLTable5}
p(y_{kd} | o_{kd} = o, l_{ko} = l, \mathbf{l}) \propto
\begin{cases}
&g_{l_{ko}}^{-y_{kd}}(y_{kd})\ \text{if $o$ exists} \\
&h_{\cdot l} g_{l}^{-y_{kd}}(y_{kd})\ \text{else if $l$ exists} \\
&\varepsilon g_{l^{new}}^{-y_{kd}}(y_{kd})\ \text{if $l = l^{new}$}
\end{cases}
\end{equation}
where $h_{\cdot l}$ is the total number tables serving time-dish $l$ and $g$ is a posterior predictive distribution of Normal, a Student's t-Distribution. $\varepsilon$ controls how likely a new time dish would be needed. Now we have finished deriving the sampling for $p(y_{kd} | t_{ji} = t, k_{jt} = k, \mathbf{y^{-kd}}, \mathbf{o^{-kd}}, \mathbf{l})$. Similar derivations can be done for $p(z_{kc} | t_{ji} = t, k_{jt} = k, \mathbf{z^{-kc}}, \mathbf{p^{-kc}}, \mathbf{q})$.

After table sampling, we need to do dish sampling (line 10 in \algref{CRFL}). The left side of \eqref{dishSampling} becomes:

\begin{multline}
\label{eq:AppCRFLDishSampling}
p(k_{jt^{new}} = k, x_{ji}, y_{kd}, z_{kc} | \mathbf{k^{-jt^{new}}}, \mathbf{y^{-kd}}, \mathbf{o^{-kd}},\\ \mathbf{l}, \mathbf{z^{-kc}}, \mathbf{p^{-kc}}, \mathbf{q}) \propto \\
\begin{cases}
&m_{\cdot k}^{-jt}p(x_{ji}|\cdots)p(y_{kd}|\cdots)p(z_{kc}|\cdots) \\
&\gamma p(x_{ji}|\cdots)p(y_{kd}|\cdots)p(z_{kc}|\cdots)
\end{cases}
\end{multline}

The differences between \eqref{AppCRFLDishSampling} and \eqref{dishSampling} are $p(y_{kd}|\cdots)$ and $p(z_{kc}|\cdots)$. Both are Infinite Gaussian Mixture Model so the likelihoods can be easily computed. We therefore have given the whole sampling process for the customer-level sampling (\eqref{AppCRFLTable}). We still need to deal with the table-level sampling.

\subsection{Table Level Sampling}
Similarly, when we do the table-level sampling (line 14 in \algref{CRFL}), the left side of \eqref{tableDishSampling} change to:

\begin{multline}
\label{eq:AppCRFLTableDishSampling}
p(k_{jt} = k, \mathbf{x_{jt}}, \mathbf{y_{kd_{jt}}}, \mathbf{z_{kc_{jt}}} | \mathbf{k^{-jt}}, \mathbf{y^{-kd_{jt}}}, \mathbf{o^{-kd_{jt}}},\\ \mathbf{l^{-ko}}, \mathbf{z^{-kc_{jt}}}, \mathbf{p^{-kc_{jt}}}, \mathbf{q^{-kp}}) \propto \\
\begin{cases}
&m_{\cdot k}^{-jt}p(\mathbf{x_{jt}}|\cdots)p(\mathbf{y_{kd_{jt}}}|\cdots)p(\mathbf{z_{kc_{jt}}}|\cdots) \\
&\gamma p(\mathbf{x_{jt}}|\cdots)p(\mathbf{y_{kd_{jt}}}|\cdots)p(\mathbf{z_{kc_{jt}}}|\cdots)
\end{cases}
\end{multline}
where $\mathbf{x_{jt}}$ is the space-customers at the table $t$, $\mathbf{y_{kd_{jt}}}$ and $\mathbf{z_{kc_{jt}}}$ are the associated time and speed customers. $\mathbf{k^{-jt}}$, $\mathbf{y^{-kd_{jt}}}$, $\mathbf{o^{-kd_{jt}}}$, $\mathbf{l^{-ko}}$, $\mathbf{z^{-kc_{jt}}}$, $\mathbf{p^{-kc_{jt}}}$, $\mathbf{q^{-kp}}$ are the rest customers and their choices of tables and dishes in three HDPs. $\cdots$ represents all the conditional variables for simplicity. $p(\mathbf{x_{jt}}|\cdots)$ is the Multinomial $f$ as in \eqref{tableDishSampling}. 

$p(\mathbf{y_{kd_{jt}}}|\cdots)$ and $p(\mathbf{z_{kc_{jt}}}|\cdots)$ are not easy to compute. However, they can be treated in the same way so we only explain how to compute $p(\mathbf{y_{kd_{jt}}}|\cdots)$ here. To fully compute $p(\mathbf{y_{kd_{jt}}}|\cdots)$ = $p(\mathbf{y_{kd_{jt}}}|k_{jt} = k, \mathbf{o^{-kd_{jt}}},\mathbf{l^{-ko}})$, one needs to consider it for every $y_{kd_{jt}} \in \mathbf{y_{kd_{jt}}}$ which is extremely expensive. This is because we deal with large datasets and there can easily be thousands, if not more, of customers in $\mathbf{y_{kd_{jt}}}$. In \eqref{AppCRFLTableDe}, we already see how $y_{kd}$'s time-restaurant preference influences the table choice of $x_{ji}$. Given a group $\mathbf{y_{kd_{jt}}}$, their collective time-restaurant preference, $p(\mathbf{y_{kd_{jt}}}|\cdots)$, will influence the dish choice of $\mathbf{x_{jt}}$. Since the distribution of individual time-restaurant preference is hard to compute analytically, we approximate it. We do a random sampling over $\mathbf{y_{kd_{jt}}}$ to approximate $p(\mathbf{y_{kd_{jt}}}|\cdots)$. This number of samples is a hyper-parameter, referred as \textit{customer selection}. For every single $y \in \mathbf{y_{kd_{jt}}}$ we can compute its probability in the same way as in \eqref{AppCRFLTable3}. So we approximate the $p(\mathbf{y_{kd_{jt}}}|\cdots)$ with the joint probability of the sampled time-customers.

\subsection{Sampling for Hyper-parameters}
\label{sec:APPHyperParameters}

A Dirichlet Process contains two parameters, a base distribution and a concentration parameter. To make THDP more robust to these parameters, we impose a prior, a Gamma distribution onto the concentration parameter $\gamma \sim \Gamma (\alpha,\varpi)$, where $\alpha$ is the shape parameter and $\varpi$ is the rate parameter. There are totally six $\alpha$s and $\varpi$s for the six DPs in THDP. They are initialized as 0.1. Then they are updated during the optimization using the method in \cite{teh_hierarchical_2006}. The update is done in every iteration in CRFL, after sampling all the other parameters. The customer selection parameter is set to 1000 across all experiments. Finally, after CRFL, the inference is done for the three distributions in \eqref{THDPTop}:
\begin{align}
\phi^s_k \sim H_s, \ \ \ \beta \sim Dirichlet (m_{\cdot 1}, m_{\cdot 2}, \cdots, m_{\cdot k}, \gamma) \\
\phi^t_l \sim H_t, \ \ \ \zeta \sim Dirichlet (h_{\cdot 1}, h_{\cdot 2}, \cdots, h_{\cdot l}, \varepsilon) \\
\phi^e_q \sim H_e, \ \ \ \rho \sim Dirichlet (a_{\cdot 1}, a_{\cdot 2}, \cdots, a_{\cdot q}, \lambda)
\end{align}
where $m_{\cdot k}$ is the total number of space-tables choosing space-dish $k$; $h_{\cdot l}$ is the total number of time-tables choosing time-dish $l$; $a_{\cdot q}$ is the total number of speed-tables choosing speed-dish $q$. $\gamma$, $\varepsilon$ and $\lambda$ are the scaling factors of $G_s$, $G_t$ and $G_e$.

\section{Simulation Guidance}
\label{sec:AppSimGuidance}
The dynamics of of one trajectory, $\bar{w}$, is:

\begin{align}
\label{eq:LDS}
    x_t^{\bar{w}} = As_t + \omega_t \ \ \ \ \omega \sim N(0, \Omega) \nonumber \\
    s_t = Bs_{t-1} + \lambda_t \ \ \ \ \lambda \sim N(0, \Lambda) \nonumber
\end{align}

Given the $U$ trajectories, from a space flow $\check{w}$, the total likelihood is:
\begin{align}
    &p(\check{w}) = \Pi_{i=1}^U p(\bar{w}_i) \ \ \ \text{where}\nonumber \\
    &p(\bar{w}_i) = \Pi_{t=2}^{T_i-1} p(x^i_t|s_t) P(s_t|s_{t-1})\ \ \, s_1 = x^i_1, s_{T} = x^i_{T_i}
\end{align}
where $A$ is an identity matrix and $\Omega$ is a known diagonal matrix. $T_{i}$ is the length of the trajectory $i$. We use homogeneous coordinates to represent both $x = [x_1, x_2, 1]^{\bold{T}}$ and $s = [s_1, s_2, 1]^{\bold{T}}$. Consequently, $A$ is a $\bold{R}^{3\times3}$ identity matrix. $\Omega$ is set to be a $\bold{R}^{3\times3}$ diagonal matrix with its non-zeros entries set to 0.001. $B$ is a $\bold{R}^{3\times3}$ transition matrix and $\Lambda$ is $\bold{R}^{3\times3}$ covariance matrix, both to be learned.

We apply Expectation-Maximization (EM) \cite{bishop_pattern_2007} to estimate parameters $B, \Lambda$ and states $S$ by maximizing the log likelihood $log\ P(\bm{u})$. Each iteration of EM consists of a E-step and a M-step. In the E-step, we fix the parameters and sample states $s$ via the posterior distribution of $x$. The posterior distribution and the expectation of complete-data likelihood are denoted as

\begin{equation}
\begin{split}
    \mathcal{L} & = E_{S|X;\hat{B},\hat{\Lambda}}(logP(S,X;B,\Lambda))\\
     &=\sum_{i}\tau_{i}E_{s^{i}|x^{i}}\{p(s^{i}, x^{i})\}
\end{split}
\end{equation}
where $\tau_{i}$ is defined as $\tau_{i} = \frac{\frac{1}{T_{i}}\sum_{t=1}^{T_{i}}p(x_{t}^{i}|s_{t}^{i})}{\sum_{i=1}^{U}\frac{1}{T^{i}}\sum_{t=1}^{T_{i}}p(x_{t}^{i}|s_{t}^{i})}$. In the M-step, we maximize the complete-data likelihood and the model parameters are updated as:

\begin{align}
    {B}^{new} = \frac{\sum_{i}\tau_{i}\sum_{t=2}^{T_{i}}P_{t,t-1}^{i}}{\sum_{i}\tau_{i}\sum_{t=2}^{T_{i}}P_{t-1,t-1}^{i}} \\
    \Lambda^{new}=\frac{\sum_{i}\tau_{i}(\sum_{t=2}^{T_{i}}P_{t,t}^{i}-B^{new}\sum_{t=2}^{T_{i}}P_{t,t-1}^{i})}{\sum_{i}\tau_{i}(T_{i}-2)}\\
    P_{t,t}^{i} = E_{s^{i}|x^{i}}(s_{t}s_{t}^{\text{T}})\\
    P_{t,t-1}^{i} = E_{s^{i}|x^{i}}(s_{t}s_{t-1}^{\text{T}})
\end{align}

During updating, we use $\Lambda = \frac{1}{2}(\Lambda + \Lambda^{\bold{T}})$ to ensure its symmetry.



\end{document}